\documentclass[a4paper,fleqn,usenatbib,useAMS]{mnras}                                  
\usepackage{graphicx}	
\usepackage{amsmath}	
\usepackage{amssymb}	
\usepackage{multicol}        
\usepackage{bm}		
\usepackage{pdflscape}	
\def\del#1{{}}

\newcommand{\expval}[1]{\left\langle #1 \right\rangle}

\newcommand{\dd}{\mathrm{d}}
\newcommand{\vel}{\upsilon}
\usepackage[T1]{fontenc}
\usepackage{ae,aecompl}

\title[On the Physics of Radio Halos in Galaxy Clusters]{On the Physics of Radio Halos in Galaxy Clusters: Scaling Relations and Luminosity Functions}
\author[F. Zandanel, C. Pfrommer and F. Prada]{
Fabio Zandanel$^{1,2}$, Christoph Pfrommer$^{3}$ and Francisco Prada$^{4,5,1}$\\
$^{1}$Instituto de Astrof\'{\i}sica de Andaluc\'{\i}a (CSIC), Glorieta de la Astronom\'{\i}a, E-18080 Granada, Spain\\
$^{2}$Now at GRAPPA Institute, University of Amsterdam, Science Park 904, 1098XH Amsterdam, Netherlands, f.zandanel@uva.nl\\
$^{3}$Heidelberg Institute for Theoretical Studies, Schloss-Wolfsbrunnenweg 35, D-69118 Heidelberg, Germany, christoph.pfrommer@h-its.org\\
$^{4}$Campus of International Excellence UAM+CSIC, Cantoblanco, E-28049 Madrid, Spain\\
$^{5}$Instituto de F\'{\i}sica Te\'orica, (UAM/CSIC), Universidad Aut\'onoma de Madrid, Cantoblanco, E-28049 Madrid, Spain}
\begin{document}

\date{Accepted 2013 November 17}

\pagerange{\pageref{firstpage}--\pageref{lastpage}} \pubyear{2016}

\maketitle

\label{firstpage}

\begin{abstract}
    The underlying physics of giant and mini radio halos in galaxy clusters
    is still an open question. We find that mini halos (such as in Perseus and
    Ophiuchus) can be explained by radio-emitting electrons that are generated
    in hadronic cosmic ray (CR) interactions with protons of the intracluster
    medium. By contrast, the hadronic model either fails to explain the extended
    emission of giant radio halos (as in Coma at low frequencies) or would
    require a flat CR profile, which can be realized through outward streaming
    and diffusion of CRs (in Coma and A2163 at 1.4~GHz).  We suggest that a
    second, leptonic component could be responsible for the missing flux in the
    outer parts of giant halos within a new hybrid scenario and we describe its
    possible observational consequences.  To study the hadronic emission
    component of the radio halo population statistically, we use a cosmological
    mock galaxy cluster catalog built from the MultiDark simulation. Because of
    the properties of CR streaming and the different scalings of the X-ray
    luminosity ($L_{\rmn{X}}$) and the Sunyaev-Zel'dovich flux ($Y$) with gas
    density, our model can simultaneously reproduce the observed bimodality of
    radio-loud and radio-quiet clusters at the same $L_{\rmn{X}}$ as well as the
    unimodal distribution of radio-halo luminosity versus $Y$; thereby
    suggesting a physical solution to this apparent contradiction. We predict
    radio halo emission down to the mass scale of galaxy groups, which
    highlights the unique prospects for low-frequency radio surveys (such as the
    LOFAR Tier 1 survey) to increase the number of detected radio halos by at
    least an order of magnitude.
\end{abstract}


\section{Introduction}
\label{sec:1}

The presence of large-scale diffuse radio synchrotron emission in clusters
  of galaxies proves the existence of relativistic electrons and magnetic fields
  permeating the intracluster medium (ICM). This diffuse cluster radio emission
  can be observationally classified into two phenomena: peripheral radio relics,
  which show irregular morphology and polarized emission and appear to trace
  merger and accretion shocks, as well as radio halos (see, e.g.,
  \citealp{2012A&ARv..20...54F}).  Radio \mbox{(mini-)}halos (RHs) are
  characterized by unpolarized radio emission, are centered on clusters and
  show a regular morphology, resembling the morphology of the thermal X-ray
  emission. However, the short cooling length of synchrotron-emitting electrons
  at GHz frequencies ($\lesssim 100$~kpc) challenges theoretical models to
  explain the large-scale radio emission that extends over several Mpcs and
  calls for an efficient in-site acceleration process of electrons.

Two principal models have been proposed to explain RHs.  In the ``hadronic
model'' the radio emitting electrons are produced in hadronic cosmic ray (CR)
proton interactions with protons of the ambient thermal ICM, requiring only a
very modest fraction of (at most) a few percent of CR-to-thermal pressure
(\citealp{1980ApJ...239L..93D,1982AJ.....87.1266V, 1999APh....12..169B,
  2000A&A...362..151D, 2001ApJ...562..233M,2001ApJ...559...59M,
  2003MNRAS.342.1009M,2003A&A...407L..73P, 2004A&A...413...17P,
  2004MNRAS.352...76P, 2007IJMPA..22..681B, 2008MNRAS.385.1211P,
  2008MNRAS.385.1242P, 2009JCAP...09..024K, 2010MNRAS.401...47D,
  2010arXiv1003.0336D, 2010arXiv1003.1133K, 2010arXiv1011.0729K,
  2011A&A...527A..99E}).  CR protons and heavier nuclei, like electrons, can be
accelerated and injected into the ICM by structure formation shocks, active
galactic nuclei (AGN) and galactic winds.  Due to their higher masses with
respect to electrons, CR protons are accelerated more efficiently to
relativistic energies and are expected to show a ratio of the spectral energy
flux of CR protons to electrons above 1 GeV of about 100, similarly to what is
observed in our Galaxy \citep{2002cra..book.....S}. Additionally, CR protons
have a radiative cooling time larger than that of the electrons by the square of
the mass ratio and therefore can accumulate in clusters for cosmological times
\citep{1996SSRv...75..279V}. In contrast, CR electrons suffer more severe energy
losses via synchrotron and inverse Compton emission at particle energies $E
\gtrsim 100$~MeV, and Coulomb losses below that energy range.

In the ``re-acceleration model'', RHs are thought to be the result of
re-acceleration of electrons through interactions with plasma waves during
powerful states of ICM turbulence, as a consequence of a cluster merger
(\citealp{1987A&A...182...21S, 1993ApJ...406..399G, 2002A&A...386..456G,
  2004MNRAS.350.1174B, 2005MNRAS.363.1173B, 2007MNRAS.378..245B,
  2010arXiv1008.0184B, 2009A&A...507..661B, 2012arXiv1211.3337D}). This,
however, requires a sufficiently long-lived CR electron population at energies
around 100~MeV which has to be continuously maintained by re-acceleration at a
rate faster than the cooling processes.  We refer the reader to
\citet{2011A&A...527A..99E} for a discussion on the strengths and weaknesses of
these two models.

RHs can be divided in two classes. Giant radio halos are typically associated
with merging clusters and have large extensions, e.g., the Coma RH has an
extension of about 2~Mpc. Radio mini-halos are associated with relaxed clusters
that harbor a cool core and typically extend over a few hundred kilo-parsecs,
e.g., the Perseus radio mini-halo has an extension of about 0.2~Mpc.  The
observed morphological similarities with the thermal X-ray emission suggests
that RHs may be of hadronic origin. In fact, cool-core clusters (CCCs) are
characterized by high thermal X-ray emissivities and ICM densities that are more
peaked in comparison to non cool-core clusters (NCCCs) that often show
signatures of cluster mergers (e.g., \citealp{2008A&A...487..431C}). This
distinct difference in the ICM density structure of CCCs and NCCCs would be
reflected in the morphology of the two observed classes of RHs.

The RH luminosity seems to be correlated to the thermal X-ray luminosity (e.g.,
\citealp{2009A&A...507..661B,2011A&A...527A..99E}). However, a large fraction of
clusters does not exhibit significant diffuse synchrotron emission at current
sensitivity limits. Stacking subsamples of luminous X-ray clusters reveals a
signal of extended diffuse radio emission that is below the radio upper limits
on individual clusters \citep{2011ApJ...740L..28B} suggesting that at least a
subset of apparently ``radio-quiet'' clusters shows a low-level diffuse
emission. Galaxy clusters with the same thermal X-ray luminosity show an
apparent bimodality with respect to their radio luminosity.  This suggests the
existence of a switch-on/switch-off mechanism that is able to change the radio
luminosity by more than one order of magnitude.  While such a mechanism could be
easily realized in the framework of the re-acceleration model
\citep{2009A&A...507..661B}, the \emph{classical} hadronic model predicts the
presence of RHs in all clusters. The failure to reproduce the observed cluster
radio-to-X-ray bimodality was one of the main criticisms against the hadronic
model.  Additionally, the classical hadronic model cannot reproduce some
spectral features observed in clusters, such as the total spectral (convex)
curvature claimed in the Coma RH or the spectral steepening observed at the
boundary of some RHs.  However, the recent report of spectral flattening with
frequency of the RH in A2256 \citep{2012A&A...543A..43V} could easily be
accommodated in the hadronic model, which naturally produces such a concave
spectrum \citep{2010MNRAS.409..449P}.  This raises the interesting question
whether such a variability among different sources that are generally classified
as ``radio halos'' signals the presence of richer underlying physics---a question
that we will address in this paper.

\cite{2011A&A...527A..99E} tried to asses these problems of the classical
hadronic model by analyzing CR transport processes within a cluster. While CR
advection with turbulently driven flows results in centrally enhanced CR
profiles, the propagation in form of CR streaming and diffusion produces a
flattening of CR profiles. Hence, different CR transport phenomena may also
account for the observed bimodality of the radio luminosity in the hadronic
model and may also explain the spectral features observed in some clusters. 
  This has been recently confirmed by \cite{2013arXiv1303.4746W}. In particular,
  by considering turbulent damping, they show that CRs can stream at
  super-Alfv{\'e}nic velocities.  Note that these phenomena were not considered in
earlier analytical works (e.g., \citealp{2004A&A...413...17P}) as well as in
previous hydrodynamic simulations (e.g., \citealp{2001ApJ...562..233M,
  2008MNRAS.385.1211P, 2010MNRAS.409..449P}).  A satisfactory theory of CR
  transport in clusters does not yet exist. However, CR streaming and diffusion
  may represent an intriguing solution for the issues of the classical hadronic
  model.

\cite{2012MNRAS.421L.112B} presented the first scaling relations
between RH luminosity and Sunyaev-Zel'dovich (SZ) flux measurements, using the
\emph{Planck} cluster catalogue. While the correlation agrees with previous
scaling measurements based on X-ray data, there is no indication for a bimodal
cluster population dividing clusters into radio-loud and radio-quiet objects at
fixed SZ flux. While the SZ flux correlates tightly with cluster mass, the X-ray
luminosity, $L_\rmn{X}$, exhibits a larger scatter. The CCCs predominantly
populate the high-$L_\rmn{X}$ tail (at any cluster temperature) and make up
approximately half of the radio-quiet objects \citep{2011A&A...527A..99E}. This
suggests that the switch-on/switch-off mechanism may not operate at fixed
$L_\rmn{X}$ but also causes an evolution of that quantity. As the cluster
relaxes after a merger, it cools and forms a denser core. Simultaneously,
$L_\rmn{X}$ is expected to {\em increase} which may simultaneously {\em
  decrease} the radio luminosity owing to the decaying turbulence that is
responsible in maintaining the radio emission in either model (that accounts for
microscopic CR transport). This has been recently confirmed by 
\cite{2013arXiv1307.3049S} and \cite{2013arXiv1306.4379C}.
 
An observational test that is able to disentangle between the hadronic and
re-acceleration models is the gamma-ray emission resulting from neutral pion
decays, a secondary product of the hadronic CR interaction with protons of the
ICM, which is not predicted by the re-acceleration model. Such observational
efforts have been undertaken in the last few years (for space-based cluster
observations in the GeV-band, see \citealt{2003ApJ...588..155R,
  2010JCAP...05..025A,2010ApJ...717L..71A,2012arXiv1207.6749H,
  2013arXiv1308.5654T,2013arXiv1308.6278H,2013arXiv1309.0197P}; for ground-based
observations in the energy band $\gtrsim100$~GeV, see
\citealt{2006ApJ...644..148P, 2008AIPC.1085..569P,
  2009arXiv0907.0727T,2009A&A...495...27A, 2009arXiv0907.3001D,
  2009arXiv0907.5000G, cangaroo_clusters, 2009ApJ...706L.275A,
  2010ApJ...710..634A, 2011arXiv1111.5544M,2012...VERITAS,2012A&A...545A.103H})
without being able to detect cluster gamma-ray emission. Current gamma-ray
limits enable us to constrain the average CR-to-thermal pressure to be less than
a couple percent, and the maximum CR acceleration efficiency at structure
formation shocks to be $<50$ per cent. Alternatively, this may indeed suggest the
presence of non-negligible CR transport processes into the outer cluster
regions.

An important step towards understanding the generating mechanism of RHs could
come from detailed RH population analyses. To date, we know of 53 clusters that
harbor RHs (\citealp{2012A&ARv..20...54F}, for an almost up-to-date list). Only
few X-ray flux-limited studies have been conducted that assess the important
question of the RH frequency in clusters \citep{1999NewA....4..141G,
  VenturiGMRT_2,2013arXiv1306.3102K}. Since the number of RHs in such X-ray
flux-limited samples is small with typically a few RHs, the conclusions on the
underlying physical mechanisms of RHs are not very robust. Fortunately, this is
expected to change thanks to the next-generation of low-frequency radio
observatories such as the Low Frequency Array (LOFAR).\footnote{www.lofar.org}
In fact, a deep cluster survey is part of the LOFAR science key projects and
expected to provide a large number of radio-emitting galaxy clusters up to
redshift $z\approx1$ \citep{2012JApA..tmp...34R}. This will hopefully permit to
clearly determine the RH phenomenology with respect to cluster properties such
as the fractions of radio-loud/quiet, non cool-core/cool-core, and
non-merging/merging clusters, and to explore the role of different parameters
like the magnetic field, the CR acceleration efficiency, and CR transport
properties.

 The main scope of this work is to account for CR transport processes in the
  classical hadronic model and to provide forecasts for future radio
  surveys. The outline is as follows. In Section~\ref{sec:2.3}, we construct a
  model for the CR proton distribution in clusters that merges results of
  hydrodynamic cluster simulations and an analytical model for microscopic CR
  transport processes. In Section~\ref{sec:3}, we model observed surface
  brightness profiles of individual RHs within the hadronic scenario and explore
  the allowable parameter space for CRs and magnetic fields. Motivated by
  immanent challenges to explain the extent of giant radio halos within the
  hadronic model, we suggest a new hybrid leptonic/hadronic scenario meant to
  unify the apparently distinct classes of giant radio halos, mini halos, and
  steep spectrum radio sources in Section~\ref{sec:discussion_hadronic}. In
  Section~\ref{sec:4}, we apply our extended hadronic model to a cosmologically
  complete mock galaxy cluster catalog built from the MultiDark $N$-body
  $\Lambda$CDM simulation in \cite{paper1}, hereafter Paper~I.  We compare the
  resulting modeled radio-to-X-ray and radio-to-SZ scaling relations to current
  observations and show how they vary for different choices of our CR and
  magnetic field parametrizations. In Section~\ref{sec:5}, we show the radio
  luminosity functions, compare them to current observational constraints, and
  provide predictions of the hadronically-generated RHs for the LOFAR cluster
  survey. Finally, in Section~\ref{sec:6}, we present our conclusions. In this
work, the cluster mass $M_{\Delta}$ and radius $R_{\Delta}$ are defined with
respect to a density that is $\Delta=200$ or $500$ times the
\emph{critical} density of the Universe. We adopt density parameters of
$\Omega_{\rmn{m}}=0.27$, $\Omega_{\Lambda}=0.73$ and today's Hubble constant of
$H_0 = 100 \, h_{70}$~km~s$^{-1}$~Mpc$^{-1}$ where $h_{70} = 0.7$.

\section{Cosmic Ray Modeling}
\label{sec:2.3}
We assume a power-law for the spectral distribution of CR protons, $f(R,p) \dd
p=C(R) p^{-\alpha} \dd p$, which is the effective one-dimensional momentum
distribution (assuming isotropy in momentum space). The spatial CR distribution
$C(R)$ within a galaxy cluster is governed by an interplay of CR advection,
streaming, and diffusion. The advection of CRs by turbulent gas motions is
dominated by the largest eddy turnover time $\tau_{\rmn{tu}}\sim L_{\rmn{tu}}/
\vel_{\rmn{tu}}$. Here, $L_{\rmn{tu}}$ denotes the turbulent injection scale
(typically of order the core radius) and $\vel_{\rmn{tu}}$ is the associated
turbulent velocity that approaches the sound speed $\vel_{\rmn{s}}$ for
transsonic turbulence after a cluster merger and relaxes to small velocities
afterwards. As a result of advection into the dense cluster atmosphere, CRs are
adiabatically compressed and experience a stratified distribution in the cluster
potential. The gradient of the CR number density leads to a net CR streaming
motion towards the cluster outskirts. Streaming CRs excite Alfv{\'e}n waves on
which they resonantly scatter \citep{1969ApJ...156..445K}. This isotropizes the
CRs' pitch angles, and thereby reduces the CR bulk speed. Balancing the growth
rate of the CR Alfv{\'e}n wave instability with the wave damping rate due to
non-linear Landau damping yields a CR streaming speed of order the Alfv{\'e}n
speed \citep{2001ApJ...553..198F}.  This increases considerably when balancing
it with the turbulent damping rate, which implies an inverse scaling with the CR
number density \citep{2013arXiv1303.4746W}.  Once CR streaming depletes the CR
number density, this causes a run-away process with a rapidly increasing
streaming speed that even surpasses the sound speed because the smaller CR
number density drives the CR Alfv{\'e}n wave instability less
efficiently. Hence, the crossing time of streaming CRs over $L_{\rmn{tu}}$ is
$\tau_{\rmn{st}}\sim \chi_B\,L_{\rmn{tu}}/ \vel_{\rmn{st}}$ with the streaming
velocity given by $\vel_{\rmn{st}}\sim \vel_{\rmn{s}}$ and $\chi_B\lesssim 1$
parametrizes the magnetic bending scale.  Magnetic bottlenecks for the
macroscopic, diffusive CR transport, are critical in lowering the microscopic
streaming velocity of CR by some finite factor. Therefore, we can define a
turbulent propagation parameter
\begin{equation}
  \label{eq:gamma_tu}
  \gamma_{\rmn{tu}}\equiv\frac{\tau_{\rmn{st}}}{\tau_{\rmn{tu}}}=
  \frac{\chi_B\,\vel_{\rmn{tu}}}{\vel_{\rmn{st}}}
\end{equation}
that indicates the relative importance of advection versus CR streaming as the
dominant CR transport mechanism. After a merger, turbulent advective transport
should dominate yielding $\gamma_{\rmn{tu}}\gg 1$, which results in centrally enhanced
CR profiles. In contrast, in a relaxed cluster, CR streaming should be the
dominant transport mechanism implying $\gamma_{\rmn{tu}}\sim1$ and producing
flat CR profiles (for a detailed discussion of these processes, see
\citealp{2011A&A...527A..99E} and \citealp{2013arXiv1303.4746W}).

We propose here to take the spectral shape of the CR distribution function from
cosmological hydrodynamical simulation of clusters \citep{2010MNRAS.409..449P},
which however did not account for CR transport. To include the latter, we
  adapt the analytical formalism of \citet{2011A&A...527A..99E}. This results in
  a model that includes the necessary CR transport physics and is able to
  predict radio and gamma-ray emission. Note that this approach is not fully
  self-consistent and points to the necessity of future hydrodynamical
  simulations to include the effect of CR streaming and diffusion on the CR
  spectrum.

\begin{figure*} 
\centering
\includegraphics[width=0.62\textwidth]{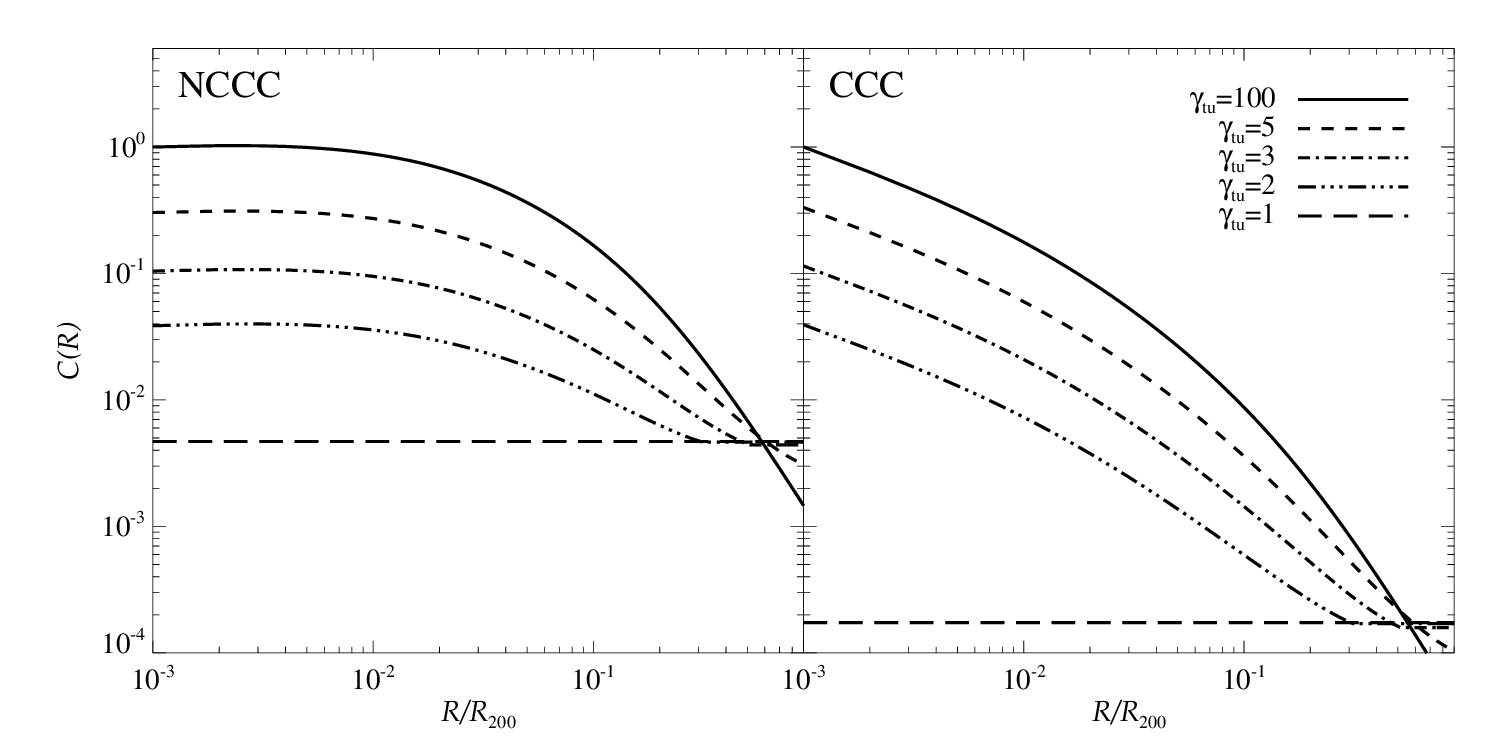}
\includegraphics[width=0.37\textwidth]{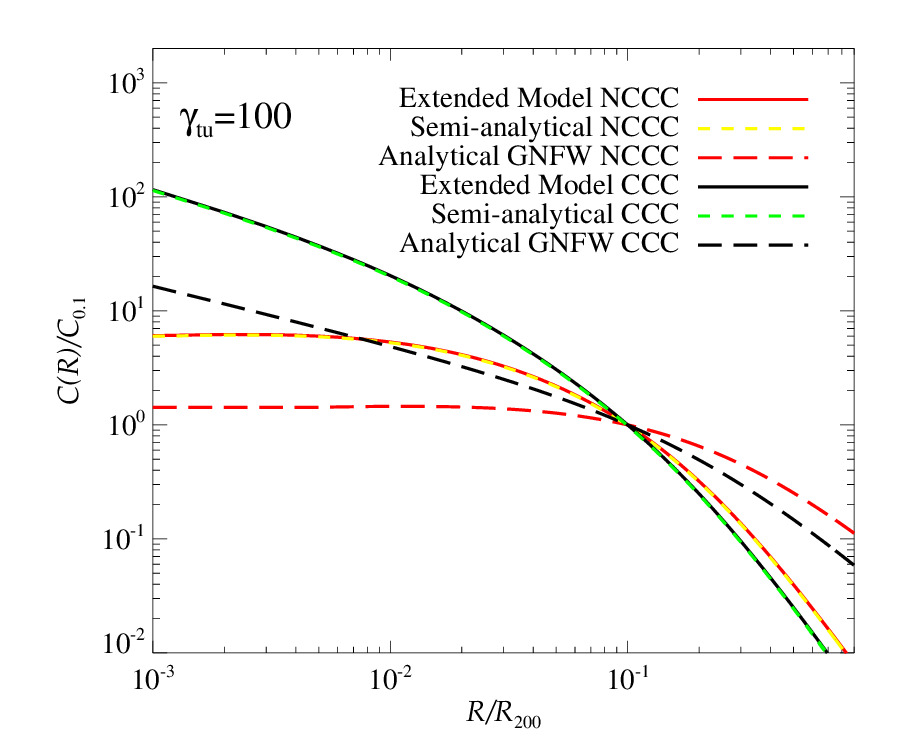}
\caption{\emph{Left panel.} We show our extended model profiles for the
  normalization of the CR distribution for the NCCC and CCC cases and for
  different values of $\gamma_{\rmn{tu}}$. We fix the CR number for the case of
  $\gamma_{\rmn{tu}}=100$ using equation~(36) of \protect\cite{2011A&A...527A..99E},
  while integrating the cluster volume within $R_{200}$, and require CR number
  conservation during CR streaming. \emph{Right panel.} We compare the
  extended model (adopting $\gamma_{\rmn{tu}}=100$) with the semi-analytical
  advection-only case (adopting our GNFW gas profiles and the outer temperature
  decrease to the simulation-derived model proposed by
  \protect\citealp{2010MNRAS.409..449P}) and with the exact analytical solution as in
  \protect\citet{2011A&A...527A..99E}, but for our GNFW profiles (adopting $\alpha=2.3$
  and $\gamma_{\rmn{tu}}=100$). Here, the CR profiles are normalized at
  $C_{0.1}=C(0.1R_{200})$.}
\label{fig:CRFinalModel}
\end{figure*}

To construct such a model, we have to generalize the approach proposed by
  \citet{2011A&A...527A..99E}, which uses a $\beta$-profile gas
  parametrisation, in order to account for different ICM gas profiles, such as
  our generalized Navarro-Frank-White (GNFW) ICM profiles derived in Paper~I
  from X-ray observations.  We also have to include the cluster mass-scaling of
the CR normalization obtained from simulations
\citep{2010MNRAS.409..449P}. While details are given in Appendix~\ref{app:B}, we
summarize below the main steps. When turbulent advection completely dominates the CR
transport, the CR normalization can be written as \citep{2011A&A...527A..99E}
\begin{equation}
C_{\rmn{adv}}(R)=C_{0} \left( \frac{P_{\rmn{th}}(R)}{P_{\rmn{th},0}} \right)^{\frac{\beta_{\rmn{CR}}}{\gamma}} = 
C_{0} \eta(R)^{\beta_{\rmn{CR}}},
\label{eq:Csimple_1}
\end{equation} 
where $P_{\rmn{th}}$ is the thermal pressure, $\beta_{\rmn{CR}}=(\alpha+2)/3$, $\gamma=5/3$, 
and we introduced the advective CR profile $\eta(R)=(P_{\rmn{th}}(R)/P_{\rmn{th},0})^{1/\gamma}$. Solving the continuity
equation for CRs, \citet{2011A&A...527A..99E} derive the CR density profile,
\begin{equation}
\rho_{\rmn{CR}}(R) = \rho_{\rmn{CR},0} \eta(R) \rmn{exp} \left( \frac{R}{R_{*}} \right) \, ,
\label{eg:rhoCR_1}
\end{equation} 
where $R_{*}=\gamma_{\rmn{\rmn{tu}}}R_{\rmn{c}}$ and $R_{\rmn{c}}$ is the characteristic 
radius, of order the core radius, at which the turbulence is supposed to be injected.
Now, we introduce the \emph{semi-analytical} mass-dependent normalization of the
CR profile of \cite{2010MNRAS.409..449P} such that
\begin{equation}
\eta(R) = \left( \frac{C_{\rmn{adv}}(R)}{C_0} \right)^{1/\beta_{\rmn{CR}}} \to
\left( \frac{C_{\rmn{extended}}(R)}{C_0} \right)^{1/\beta_{\rmn{CR}}} \, ,
\label{eq:eta}
\end{equation} 
which effectively redefines $C_{\rmn{adv}}(R)$ by that of our extended model, i.e.,
\begin{equation}
C_{\rmn{extended}}(R) =  \tilde{C}(R)\, \frac{\rho_{\rmn{gas}}(R)}{m_\rmn{p}} \frac{T(R)}{T_0}.
\label{eq:Cf}
\end{equation} 
Here, $\rho_{\rmn{gas}}$ is the ICM gas density and $\tilde{C}(R)$ is the
normalization of the CR profile of equation~(22) of
\cite{2010MNRAS.409..449P}. We additionally account for the temperature decline
toward the cluster periphery, $T(R)$, given by the fit to the universal
temperature profile obtained from cosmological hydrodynamical simulations
\citep{2007MNRAS.378..385P,2010MNRAS.409..449P} and deep {\em Chandra} X-ray
observations \citep{2005ApJ...628..655V}. Eventually, the CR profile in our
extended model is given by
\begin{equation}
C(R)=C_{0}\left(\frac{\rho_{\rmn{CR}}(R)}{\rho_{\rmn{CR},0}}\right)^{\beta_{\rmn{CR}}},
\label{eq:Cf2}
\end{equation} 
which is valid within $R_{\pm}$ (equation~\ref{eq:Rpm}), with
$\rho_{\rmn{CR}}$ defined by equation~(\ref{eg:rhoCR_1}) where
$C_{\rmn{extended}}$ enters through our redefinition of $\eta$, and $C(R) =
C(R_{\pm})$ for $R > R_{+}$ and $R < R_{-}$, respectively.

The last step is to generalize the case of one CR population with a single
spectral index $\alpha$ to include the spectral curvature as suggested by
\cite{2010MNRAS.409..449P}. They model the CR spectrum with three different
power-law CR populations with spectral indices of
$\alpha_{i}=(2.15,2.3,2.55)$. Our formalism can be easily extended to
  account for multiple CR populations by extending the terms with a single
  $\alpha$ to sums over the three spectral indices
  (\citealp{2010MNRAS.409..449P}, see Appendix~\ref{app:A}). However,
  introducing a sum over $\alpha_{i}$ in equation~(\ref{eq:Csimple_1}) would
  make it impossible to solve analytically for $\eta(R)$ in
  equation~(\ref{eq:eta}). For simplicity, we decided to only use $\alpha =
2.3$ in this last case.\footnote{We checked that the choice of $\alpha$ in
  equation~(\ref{eq:Csimple_1}) has only a minor effect on the results. Varying
  $\alpha$ within $2.15-2.55$ yields a similar radial shape and normalization
  within 0.5 per cent.} For the highly turbulent cases, i.e., for
$\gamma_{\rmn{tu}}=100$ (1000), we recover the radial shape and normalization of
the semi-analytical model of \cite{2010MNRAS.409..449P} within 1 per cent (0.1
per cent).

Summarizing, our \emph{extended} model for the CR distribution function, has the
following properties: it accounts for (i) the X-ray-inferred gas
profiles and cluster-mass scaling of the gas fraction (see Paper~I), in addition to the
universal temperature drop in the outskirts of clusters, (ii) a cluster-mass
dependent CR normalization and universal CR spectrum as derived from
cosmological hydrodynamical simulations, (iii) an effective parametrization of
active CR transport processes, including CR streaming and diffusion, which
allows us to explore different turbulent states of the clusters in our mock
cluster catalog.

In the left two panels of Fig.~\ref{fig:CRFinalModel}, we show our extended CR 
normalization for the GNFW gas profile in the NCCC and CCC cases (see Paper I) 
and for different values of $\gamma_{\rmn{tu}}$.  
As expected, when CR streaming is the dominant CR
transport mechanism, i.e., for negligible advective turbulent transport or
equivalently, $\gamma_{\rmn{tu}}\sim1$, the spatial CR profiles are flattened
irrespective of the cluster state. While turbulence in NCCCs could be
injected by a merging (sub-)cluster, in the case of CCCs, the interaction of the
AGN jet or radio lobe with the ambient ICM could be the source of turbulence.

In the right panel of Fig.~\ref{fig:CRFinalModel}, we compare our extended model
profile with the semi-analytical {\em advection-only case} (adopting our GNFW
gas profiles and the outer temperature decrease to the model proposed by
\citealp{2010MNRAS.409..449P}) and with the exact analytical solution as in
\citet{2011A&A...527A..99E}, but for our GNFW profiles (see Appendix~\ref{app:B}
for details).  The profiles are normalized at $0.1 R_{200}$. In the case of
dominant advective CR transport, our extended model is an excellent match to the
semi-analytical model derived from cosmological cluster simulations
\citep{2010MNRAS.409..449P}.  The main differences between our extended model
(and the semi-analytical model) on the one side and the analytical solution on
the other side is the inclusion of the simulation-based ``reference'' profile
$\tilde{C}$ for the advection-only case and the universally observed temperature
drop towards the outskirts of clusters. Note that the profiles in our extended
model are generally more centrally peaked in comparison to the analytical GNFW
case, which is due to the enhanced radiative cooling in the
\citet{2010MNRAS.409..449P} simulations that did not account for AGN
feedback. Thanks to the flexible parametrization in our model, this can be
easily counteracted by changing $\gamma_{\rmn{tu}}$ and $ \alpha_{\rmn{B}}$
(representing the magnetic field radial decline, see next section), however, at
the expense that these parameters are now degenerate with our assumptions on the
CR profile in the advection-dominated regime and other possible effects that we
are not considering, such as cluster asphericity.

\section{Radio Surface Brightness Modeling}
\label{sec:3}

In this section, we apply our model to reproduce the emission characteristics of
four well-observed RHs.  We provide the synchrotron emissivity $j_{\nu}$,
  at frequency $\nu$ and per steradian, in Appendix~\ref{app:A}.  The radio
  surface brightness $S_{\nu}(R_{\perp})$ (in the small-angle approximation) and
  luminosity $L_{\nu}$, at a given frequency $\nu$, are given by
\begin{eqnarray}
S_\nu(R_{\perp}) &=& 2 \int_{R_{\perp}}^{\infty} j_{\nu}(R) \frac{R}{\sqrt{R^{2}-R_{\perp}^{2}}} \rmn{d}R, \label{eq:surf} \\
L_{\nu}  &=&  4 \pi \int \dd V j_\nu(R).
\label{eq:lum}
\end{eqnarray}
The flux is given by $F_{\nu}=L_{\nu}/(4\pi D^{2})$ where $D$ is the luminosity
distance to the object. Note that we do not convolve $S_\nu$ with the instrumental
point spread function unless specified.

For the purpose of this section, we adopt the measured gas and temperature
profiles derived from X-ray observations of each cluster.  Our extended model
includes an overall normalization $g_{\rmn{CR}}$ of the CR distribution function
and the hadronically-induced non-thermal emission (Appendix~\ref{app:A}). Note
that this parameter can be interpreted as a functional that depends on the
\emph{maximum CR acceleration efficiency}, $g(\zeta_{\rmn{p,max}})$, but
\emph{only} for $\gamma_{\rmn{tu}}\gtrsim100$ \citep{2010MNRAS.409..449P}. We
will additionally study the CR-to-thermal pressure $X_{\rmn{CR}}=
P_{\rmn{CR}}/P_{\rmn{th}}$, where the CR pressure is given by
\begin{equation}
  \label{eq:PCR}
  P_{\rmn{CR}}=\frac{g_{\rmn{CR}} C m_{\rmn{p}} c^{2}}{6}
  \sum_{i=1}^{3} \Delta_{i} \mathcal{B}_{1/(1+q^2)} \left(
    \frac{\alpha_{i}-2}{2},\frac{3-\alpha_{i}}{2} \right).
\end{equation}
Here, $c$ is the speed of light, $\mathcal{B}_x(a,b)$ denotes the incomplete beta
function, $q=0.8$ is the low-momentum cutoff of the CR distribution, and the
normalization factors of the individual CR populations are given by $\Delta_{i}
= (0.767, 0.143, 0.0975)$ \citep[][see also
Appendix~\ref{app:A}]{2010MNRAS.409..449P}.

We assume a scaling of the magnetic field with gas density that is given by
\begin{equation}
B(R) = B_0\,\left(\frac{\rho_{\rmn{gas}}(R)}{\rho_{\rmn{gas},0}}\right)^{ \alpha_{\rmn{B}}},
\label{eq:B}
\end{equation}
where $B_0$ is the central magnetic field and $\alpha_{\rmn{B}}$ describes the
declining rate of the magnetic field strength toward the cluster outskirts. Such
a parametrization is suggested by cosmological simulations \citep{2008A&A...482L..13D} 
as well as Faraday rotation measurements 
\citep[][and references therein]{2010A&A...513A..30B, 2011A&A...529A..13K}.

For our study, we choose the giant radio halos of Coma
\citep{1997A&A...321...55D} and Abell~2163 \citep{2001A&A...373..106F,
  2009A&A...499..679M}, both in merging NCCCs, and the radio mini-halos of
Perseus \citep{1990MNRAS.246..477P} and Ophiuchus \citep{2009A&A...499..371G,
  2009A&A...499..679M}, both in relaxed CCCs. The radio emission of these
clusters is representative of a wide class of RHs.  Additionally, Perseus,
Ophiuchus and Coma are among the most promising clusters for gamma-ray
observations \citep{2010MNRAS.409..449P,2011arXiv1105.3240P}. 
We use X-ray-inferred gas densities $\rho_{\rmn{gas}}$ and temperatures 
for Coma \citep{1992A&A...259L..31B}, for A2163 and Ophiuchus 
\citep{2002ApJ...567..716R}, and for Perseus \citep{2003ApJ...590..225C}. 
In Table~\ref{tab:RadioHalos}, we summarize the main characteristics of these RHs.
 
\begin{figure*}
\centering
\includegraphics[width=5.5cm,height=5.5cm,keepaspectratio]{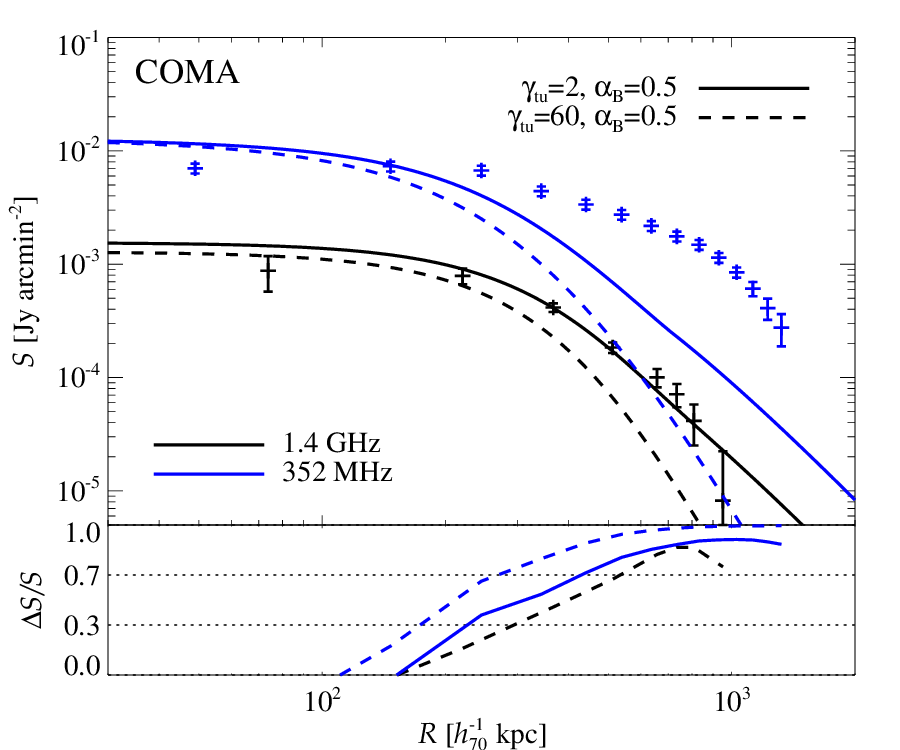}
\includegraphics[width=5.6cm,height=5.6cm,keepaspectratio]{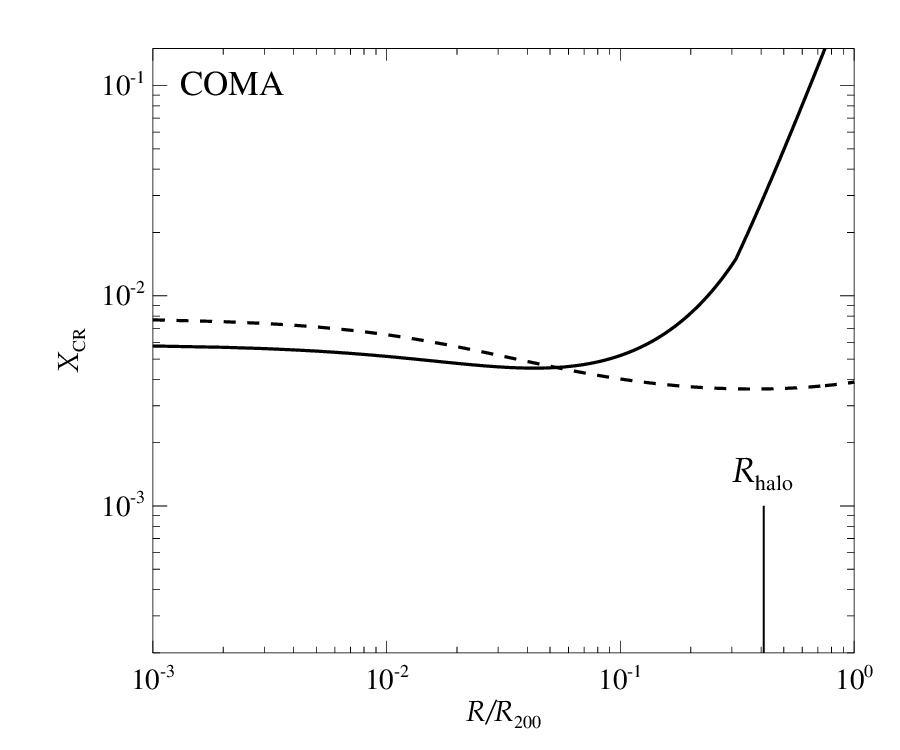}
\includegraphics[width=5cm,height=4.6cm]{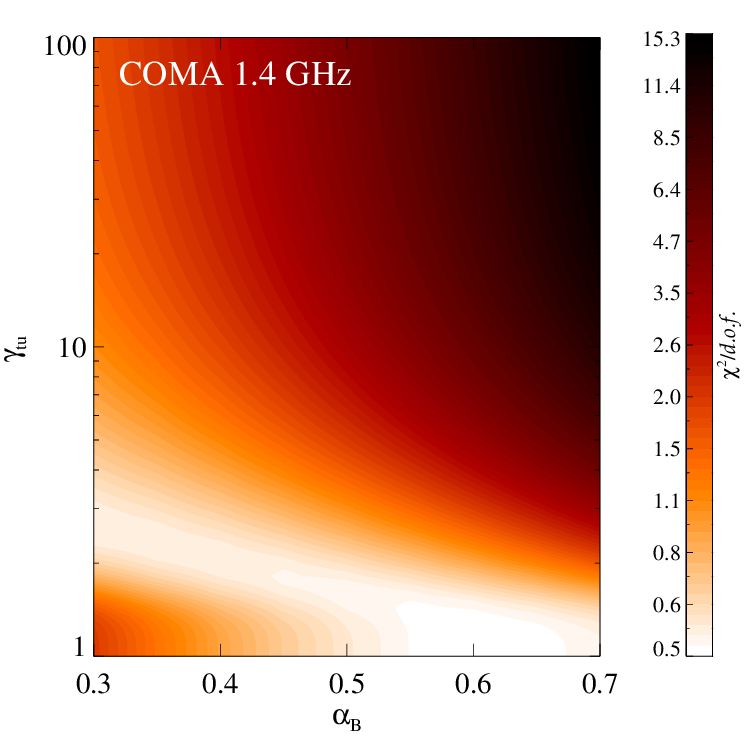}
\includegraphics[width=5.5cm,height=5.5cm,keepaspectratio]{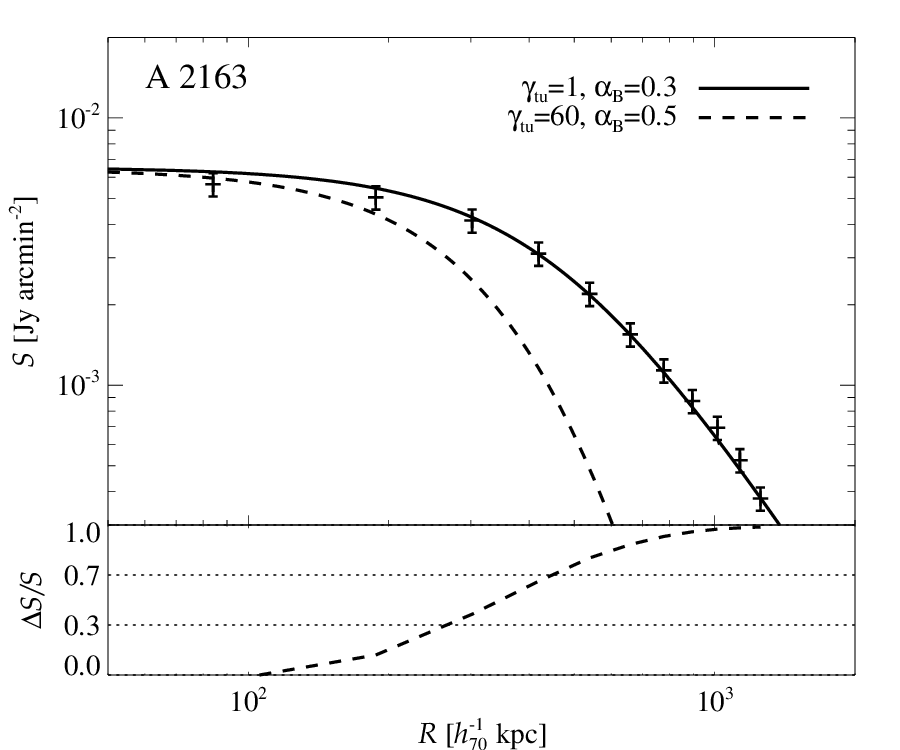}
\includegraphics[width=5.6cm,height=5.6cm,keepaspectratio]{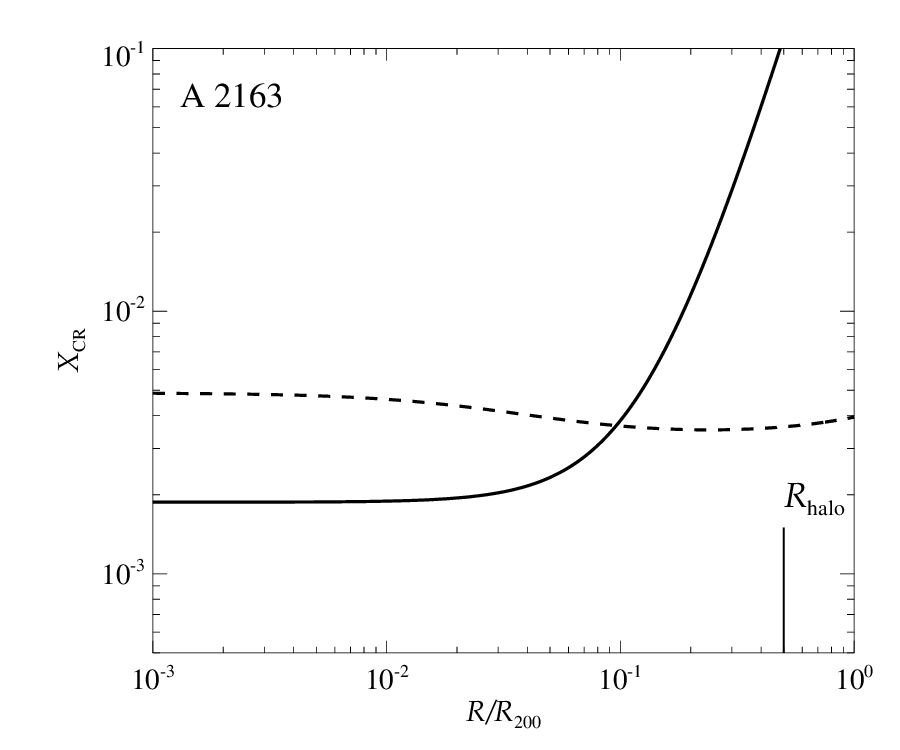}
\includegraphics[width=5cm,height=4.6cm]{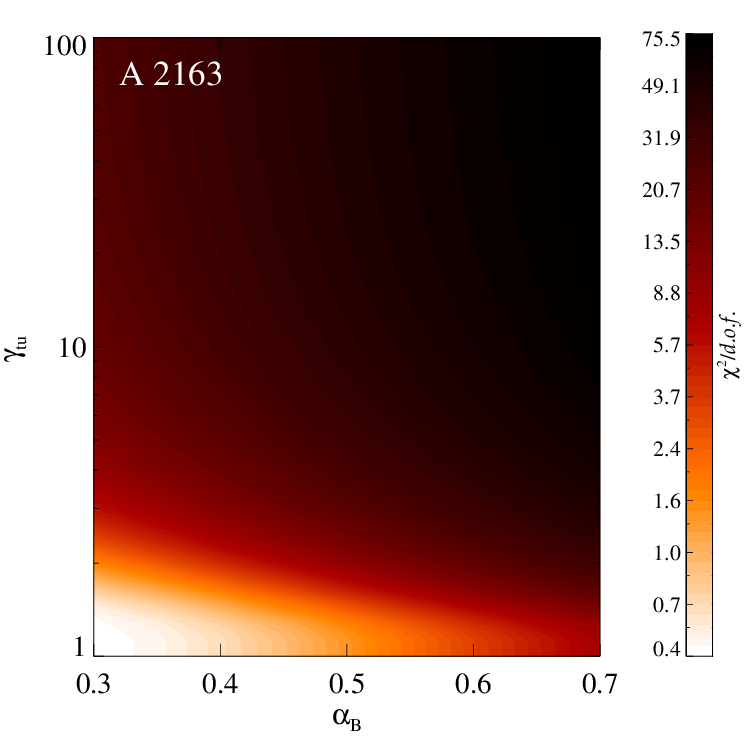}
\includegraphics[width=5.6cm,height=5.6cm,keepaspectratio]{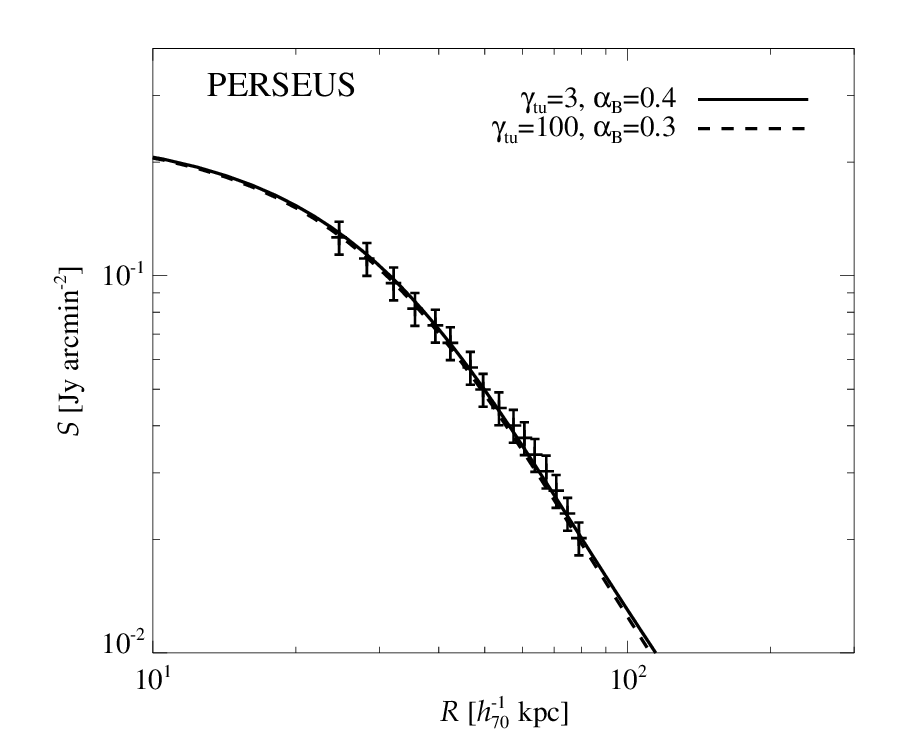}
\includegraphics[width=5.6cm,height=5.6cm,keepaspectratio]{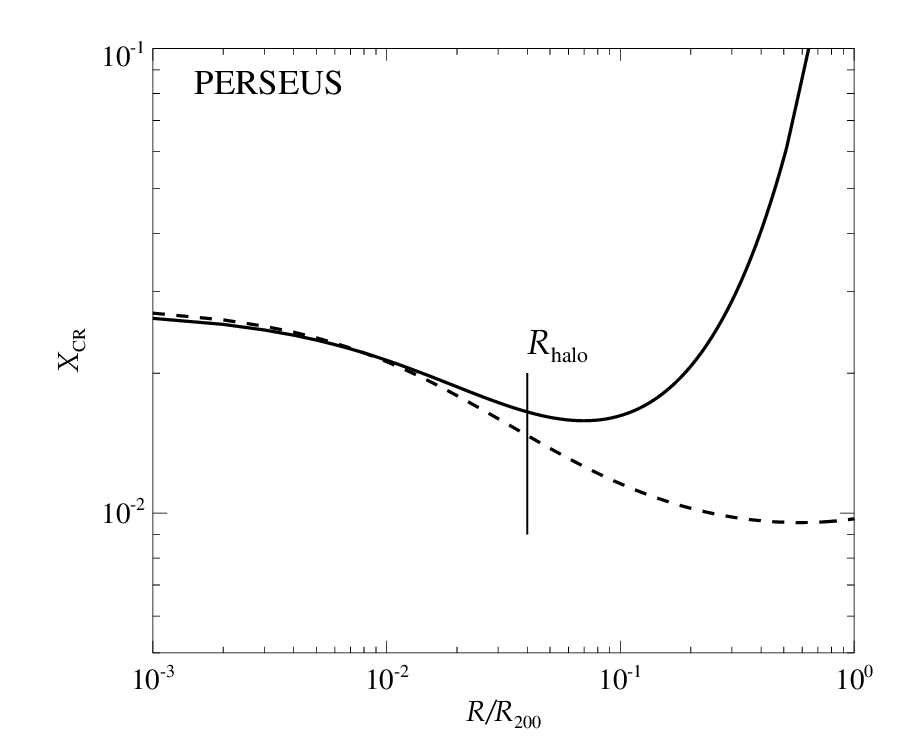}
\includegraphics[width=5cm,height=4.6cm]{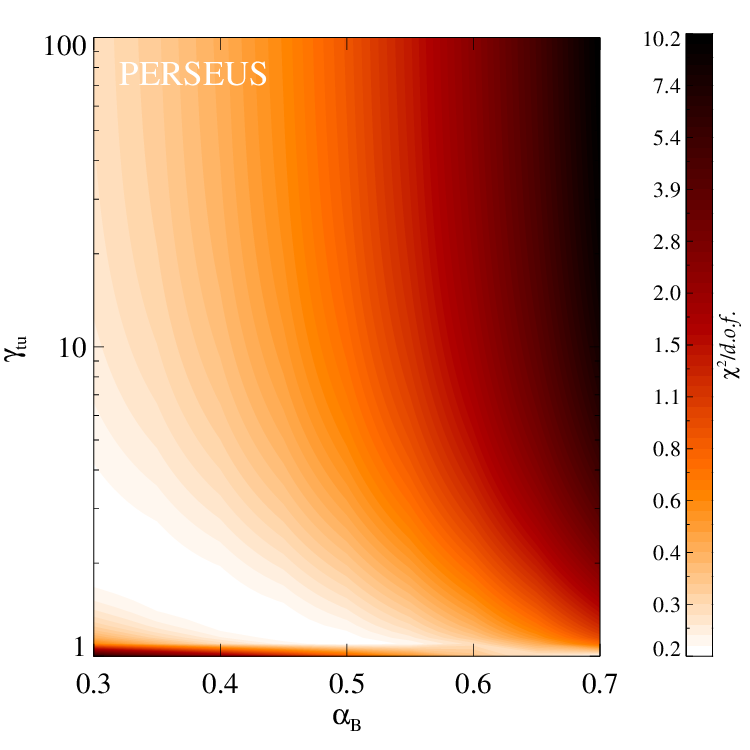}
\includegraphics[width=5.6cm,height=5.6cm,keepaspectratio]{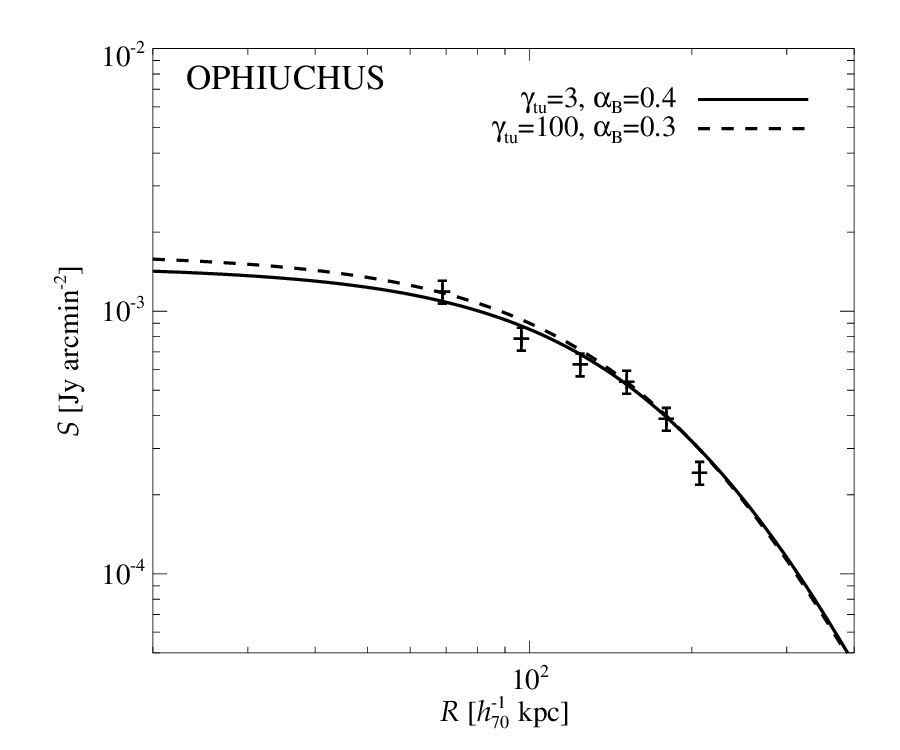}
\includegraphics[width=5.6cm,height=5.6cm,keepaspectratio]{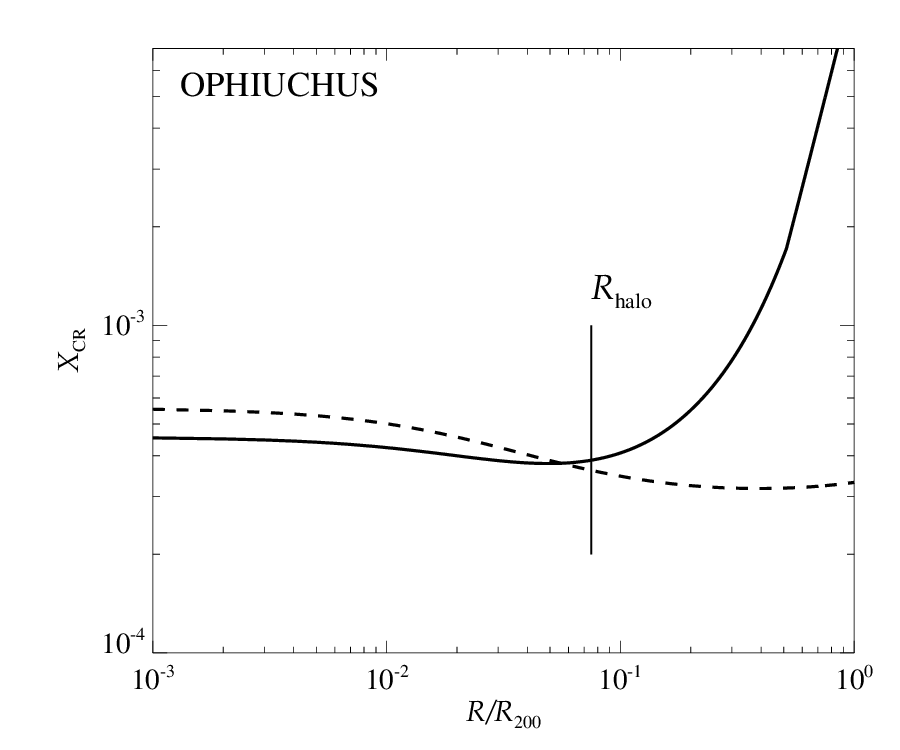}
\includegraphics[width=5cm,height=4.6cm]{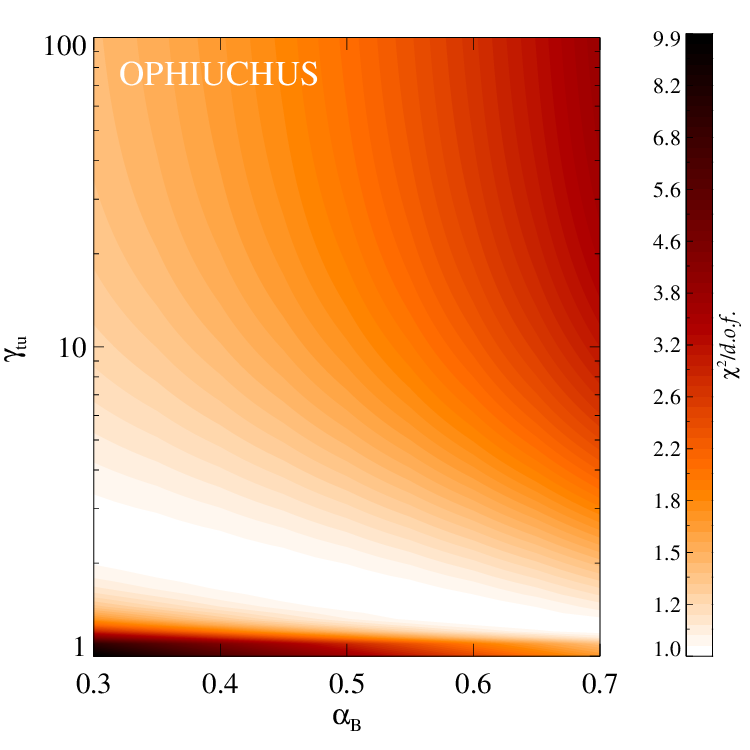}
\caption{Surface brightness modeling of the RHs in Coma, Abell~2163, Perseus and
  Ophiuchus. The left and middle panels show the RHs' azimuthally averaged
  surface brightness profiles and the corresponding CR-to-thermal pressure
  profiles $X_{\rmn{CR}}(r)$, respectively. Representative hadronic model
  parameters that fit the data well (solid) are compared to parameter choices
  that will be used in the second part of the paper (dashed, Sects.~\ref{sec:4}
  and \ref{sec:5}), which addresses RH statistics. While radio mini-halos can be
  fit by either set of parameters, for the latter choice of parameters, the
  hadronic model is not able to explain the emission in the outer parts of giant
  radio halos and would need a secondary, leptonic component (see text for
  details). This is exemplified in the lower left panels for Coma and Abell~2163
  that show the fraction of missing surface brightness for these parameter
  choices. In the middle panels, we additionally mark the RHs' radial extension
  by a vertical line. The panels on the right show the reduced-$\chi^2$ values
  of our model fits to the data in the $\gamma_{\rmn{tu}}-\alpha_{B}$ parameter
  space.  Regions of parameter space with reduced-$\chi^2$ values substantially
  larger than unity are excluded by the data while values much smaller than
  unity may point to an overestimate of the uncertainty intervals. Note that
  different parameter values that yield almost the same surface brightness
  profiles may result in very different $X_{\rmn{CR}}$ profiles. In the case of
  Abell~2163 and Ophiuchus, we adopt a 10 per cent uncertainty range instead of
  the errors reported by \citet{2009A&A...499..679M} to account for additional
  systematic uncertainties, e.g., residual point source contamination. For
  Perseus, we show the mini-halo data only for the range that is unaffected by
  residual point sources \citep{1990MNRAS.246..477P} and adopt a 10 per cent
  uncertainty budget.}
\label{fig:SBmodeling}
\end{figure*}

To assess the ability of our extended hadronic model to fit the observed surface
brightness profiles, we scan our physically motivated parameter space. The
  free parameters are the magnetic field (parametrized by $B_0$ and $
  \alpha_{\rmn{B}}$), the turbulent CR propagation parameter
  ($\gamma_{\rmn{tu}}$) and the CR acceleration efficiency. Generally, the
normalization of the magnetic profile ($B_0$) and the CR acceleration efficiency
function ($g_{\rmn{CR}}$) determine the overall normalization of the
emission. The radial decline of the magnetic field ($ \alpha_{\rmn{B}}$) and
$\gamma_{\rmn{tu}}$, both determine the shape of the radio profile and, hence,
are also degenerate.  By scanning the allowed parameter space and asserting
Bayesian priors that rely on observational constraints and theoretical
considerations about likely parameter combinations for certain classes (mini
halos versus giant halos), we will draw conclusions on the applicability of the
hadronic model for RHs.  In Fig.~\ref{fig:SBmodeling}, we show the surface
brightness and CR-to-thermal pressure profiles of each cluster together with the
allowed $\gamma_{\rmn{tu}}-\alpha_{B}$ parameter space.  All these clusters are
modeled at 1.4~GHz and within $R_{200}$, unless differently specified.

\begin{table} 
\begin{center}
\caption{Radio-halo and mini-halo characteristics.}
\medskip
\begin{tabular}{lcrrc}
\hline
\phantom{\Big|}
 cluster & $z$ & $D$~~ & $L_{1.4~\rmn{GHz}}$ & references \\
\hline \\[-0.5em]
Coma           & $0.023$ & $101$ & $0.72$  &  [1]   \\
A2163         & $0.203$ & $962$ & $15.36$  &  [2]  \\
\hline \\[-0.5em]
Perseus        & $0.018$ & $78$   & $4.40$ &  [3]  \\
Ophiuchucs     & $0.028$ & $121$  & $0.19$  &  [2] \\[0.5em]
\hline
\end{tabular}
\label{tab:RadioHalos}
\end{center}
\footnotesize{Note. Top two rows correspond to giant radio halos, while the
  bottom two rows are radio mini-halos.  $D$ is the luminosity distance in units
  of $h_{70}^{-1}$~Mpc and $L_{1.4~\rmn{GHz}}$ is the observed radio luminosity
  at 1.4~GHz in units of $10^{31}$~$h_{70}^{-2}$~erg~s$^{-1}$~Hz$^{-1}$.
  References: [1] \cite{1997A&A...321...55D} [2] \cite{2009A&A...499..679M} [3]
  \cite{1990MNRAS.246..477P}.}
\end{table}

\del{$M_{200}$ and $R_{200}$ are taken from \cite{2002ApJ...567..716R}. }

\subsection{The Coma radio halo}

The giant radio halo in Coma has a morphology remarkably similar to the extended
X-ray thermal bremsstrahlung emission, although the radio emission declines more
slowly towards the cluster outskirts \citep{1992A&A...259L..31B,
  1997A&A...321...55D}. The morphology is non-spherical, showing an elongation
in the East-West direction.  The full-width half maximum (FWHM) of the radio
beam is $0.156$~deg \citep{1997A&A...321...55D}, almost two orders of magnitude
larger than the angular resolution of the X-ray observation of
\cite{1992A&A...259L..31B}.\footnote{The apparent displacement of the radio and
  X-ray peak of about $0.05$~deg is well within the angular resolution of the
  radio observation and hence negligible for the modeling.}  Thus, we apply a
Gaussian smoothing to our theoretical surface brightness of
equation~(\ref{eq:surf}) with $\sigma_{\rmn{smoothing}} =
FWHM_{\rmn{radio}}/2.355$.

We investigate different values for $\alpha_{\rmn{B}}\in [0.3,0.7]$ and
$\gamma_{\rmn{tu}}\in [1,100]$. First, we determine the CR number for
$\gamma_{\rmn{tu}}=100$ using equation~(36) of \cite{2011A&A...527A..99E} while
integrating the cluster volume within $R_{200}$. Then, we require CR number
conservation during CR streaming (for CR energies $E>$~GeV where Coulomb cooling
is negligible for CR protons), which is realized in our model by lowering the
values of $\gamma_{\rmn{tu}}$. Fixing the central magnetic field
$B_{0}=5$~$\mu$G \citep{2010A&A...513A..30B}, we use $g_{\rmn{CR}}$ as
normalization factor to match the radio observations.  The study of the
$\gamma_{\rmn{tu}}-\alpha_{B}$ parameter space shown in
Fig.~\ref{fig:SBmodeling} (top right panel) demonstrates the necessity of low
values of $\gamma_{\rmn{tu}}$ to match the data, i.e., very flat CR profiles. An
example of such a good match to the data is obtained for $\gamma_{\rmn{tu}}=2$
and $\alpha_{\rmn{B}}=0.5$ (top left panel). Values as high as
$\gamma_{\rmn{tu}} \approx 4$ still provide an acceptable fit, however, at the expense
of a shallower decline of the magnetic field profile (smaller $ \alpha_{\rmn{B}}$) as a
function of cluster-centric radius. With such values, we can recover the shape
of the radio surface brightness as well as the total radio luminosity with a
maximal relative deviation of about 25 per cent.

The gamma-ray flux (Appendix~\ref{app:C}) within $R_{200}$ for the parameter
combination $\gamma_{\rmn{tu}} = 2$ and $\alpha_{\rmn{B}}=0.5$ and for energies
above 100~MeV (100~GeV) is $F_{\gamma} = 2.4 \times 10^{-9}$ ($8.7 \times
10^{-13}$) cm$^{-2}$~s$^{-1}$.  We note that our modeled gamma-ray flux
  $F_{\gamma} (>500~\rmn{MeV}) = 6.9 \times10^{-10}$~cm$^{-2}$~s$^{-1}$ formally
  violates the upper limit recently set with the \emph{Fermi}-LAT data of
  $F_{\gamma, UL} (>500~\rmn{MeV}) = 4 \times 10^{-10}$~cm$^{-2}$~s$^{-1}$
  \citep{2013arXiv1308.5654T}.  However, this upper limit has been obtained for
  the {\em advection-only} case \citep{2010MNRAS.409..449P}, which is
  significantly more peaked than the streaming-dominated $\gamma_{\rmn{tu}} = 2$
  case considered here and , thus, it is not directly applicable.  Note also
that for slightly higher values of $\gamma_{\rmn{tu}}$, i.e., a more centrally
concentrated CR distribution, the radio and gamma-ray yield would be increased
(assuming CR number conservation). However, in order to match the observed radio
synchrotron profiles, we have to decrease the CR normalization (parametrized by
$g_{\rmn{CR}}$). This causes the associated gamma-ray flux also to be reduced to
a level that is low enough to easily circumvent the gamma-ray constraints.  
  E.g., for the parameter combination $\gamma_{\rmn{tu}} = 3$ and
  $\alpha_{\rmn{B}}=0.4$, we obtain $F_{\gamma} = 1.3 \times 10^{-9}, 3.9 \times
  10^{-10}$, and $4.9 \times 10^{-13}$~cm$^{-2}$~s$^{-1}$ for energies above
  100~MeV, 500~MeV, and 100~GeV, respectively.  In principle, CR streaming
should cause the CR spectrum to steepen \citep{2013arXiv1303.4746W}.  This may
then considerably weaken these constraints as a result of the convex spectral
curvature since the gamma-ray emission probes the high-energy tail of the CR
distribution that is suppressed in this picture in comparison to the
lower-energy protons that the radio emission is sensitive to \citep[see][for an
extended discussion of this point]{2011arXiv1111.5544M}.

However, such low values of $\gamma_{\rmn{tu}}$ challenge the picture
that only clusters that are characterized by a highly turbulent state can host
giant radio halos. For illustration, in Fig.~\ref{fig:SBmodeling}, we
additionally show the radio surface brightness for $\gamma_{\rmn{tu}} = 60$ and
$\alpha_{\rmn{B}}=0.5$; the corresponding gamma-ray flux above 100~MeV 
(100~GeV) is $F_{\gamma} = 5.4 \times 10^{-10}$ ($1.9 \times
10^{-13}$) cm$^{-2}$~s$^{-1}$. 
Clearly, the hadronic model is not able to explain the
emission in the outer halo parts and would need a secondary component to fill in
the ``missing'' hadronic radio emission. This is exemplified in the lower plot of
the top left panel of Fig.~\ref{fig:SBmodeling}, which shows the fraction of
missing surface brightness as a function of radius and accumulates to a total
missing power of about 35 per cent.

  The much more extended RH profile at 352~MHz represents a serious challenge
  for our extended hadronic model \citep{2012arXiv1207.3025B}. We complement our
  RH modeling at high frequencies (1.4 GHz) with modeling of the new data at
  352~MHz \citep{2011MNRAS.412....2B}. To this end, we use a {\em novel}
  $352$~MHz surface brightness profile that was corrected for residual
  point-source contamination by applying the multi-resolution filtering
  technique described in \citet{2002PASP..114..427R} as well as adopting the
  X-ray center for the RH profile (Rudnick, priv.{\ }comm.). The resulting
  profile (shown in blue in Fig.~\ref{fig:SBmodeling}) declines at a slightly
  faster rate towards the outskirts than the profile used by
  \citet{2012arXiv1207.3025B}. More importantly, there is considerable azimuthal
  variation in the halo profile (see Fig.{\ }4 of \citealp{2011MNRAS.412....2B}
  and also our discussion about RH asphericity in the next section), which would
  eventually have to be modeled through hydrodynamical cluster simulations but
  which is beyond the scope of this work.

As shown by the two model realizations in Fig.~\ref{fig:SBmodeling}, the
(extended) hadronic model cannot account for the total emission at $352$~MHz for
any value in the ($\gamma_{\rmn{tu}}$, $\alpha_{\rmn{B}}$) parameter space; in
agreement with the findings of \citet{2012arXiv1207.3025B}. At the same time,
our analysis at 1.4~GHz confirms the result by the \citet{2012...VERITAS} who
also conclude that the hadronic model for the Coma RH is a viable explanation
for magnetic field estimates inferred by Faraday rotation measure studies
\citep{2010A&A...513A..30B} and is not challenged by {\em Fermi} upper limits on
the gamma-ray emission. However, this model agreement is bought at the expense
of flat CR profiles (i.e., {\em low} $\gamma_{\rmn{tu}}$ values) that are
contrary to the expectation of turbulent clusters to host giant radio halos
(i.e., {\em high} $\gamma_{\rmn{tu}}$ values) as proposed by
\citet{2011A&A...527A..99E}. Note that \cite{2013arXiv1303.4746W} arrive at
  a different conclusion and find that the increase of turbulence promotes
  outward streaming more than inward advection, thus enabling flat CR
  distributions in turbulent clusters. However, this does not help in the case
  of the 352~MHz data, where, as discussed above, not even a flat CR profile
  would suffice to explain the observed emission within the hadronic scenario.
These finding hint at the necessity of a second, leptonic component (within the
general framework of the hadronic model) that fills in the patchier emission in
the peripheral, low-surface brightness regions of the halo
\citep{2008MNRAS.385.1211P}, in particular at low frequencies \cite[see Fig.~3
of][]{2011MNRAS.412....2B}. We will return to this point in
Section~\ref{sec:discussion_hadronic}.

\subsection{The radio halo in Abell~2163}

The morphology of the giant radio halo in Abell~2163 is also closely correlated
to the cluster's thermal X-ray structure. As in Coma, the radio emission
declines towards the cluster outskirts at a slower rate in comparison to the
thermal X-ray emission \citep{2001A&A...373..106F}. The morphological appearance
is non-spherical, with an elongation in the East-West direction. We use the
surface brightness map provided by \citet{2009A&A...499..679M} for which the
synthesized radio beam can be approximated by a circular Gaussian with
$FWHM_{\rmn{radio}}=62\arcsec$.  Again, $FWHM_{\rmn{radio}}$ is larger than the
angular resolution of the \emph{ROSAT} observation and the corresponding gas
density profile. Converted to physical scale, $\sigma_{\rmn{smoothing}}$ is of
the order of that of Coma because of the larger distance of Abell~2163. Hence,
we also apply Gaussian smoothing.

We follow the same procedure as in Coma, and adopt a central magnetic field
strength of $B_{0}=5~\mu$G. Similar to the case of Coma (in fact even more
extremely) only very low values of $\gamma_{\rmn{tu}}$ provide a good match to
the data.  In Fig.~\ref{fig:SBmodeling}, we show the case of
$\gamma_{\rmn{tu}}=1$ and $ \alpha_{\rmn{B}}=0.3$, i.e., the flattest possible surface
brightness. With this choice of parameters, we recover the emission shape and
the total luminosity within about 15 per cent.  The corresponding gamma-ray flux
within $R_{200}$ is $F_{\gamma} (>100~\rmn{MeV}) = 4.2
\times10^{-10}$~cm$^{-2}$~s$^{-1}$, about two orders of magnitude lower than the
upper limit obtained by \emph{Fermi}-LAT \citep{2010ApJ...717L..71A}, and
$F_{\gamma} (>100~\rmn{GeV}) =1.5 \times10^{-13}$~cm$^{-2}$~s$^{-1}$.

As for Coma, in Fig.~\ref{fig:SBmodeling}, we show the model surface brightness
for the parameter combination $\gamma_{\rmn{tu}} = 60$ and
$\alpha_{\rmn{B}}=0.5$. The corresponding gamma-ray flux above 100~MeV
  (100~GeV) is $F_{\gamma} = 5.9 \times 10^{-11}$ ($2.2 \times 10^{-14}$)
  cm$^{-2}$~s$^{-1}$. The lower panel shows the fraction of missing surface
brightness of our model to explain the data as a function of radius. That
fraction accumulates to a total missing power of about 80 per cent for the giant
radio halo.

\subsection{The Perseus radio mini-halo}

The diffuse radio emission in Perseus is the best known example of a radio
mini-halo \citep{1990MNRAS.246..477P}\footnote{We make use of the
  \citet{1990MNRAS.246..477P} data instead of \citet{Sijbring1993} as the latter
  may be affected by residual point-source contamination.}  and Perseus itself
is among the best studied clusters in X-rays
(e.g.,~\citealp{2003ApJ...590..225C,2006MNRAS.366..417F,2011arXiv1105.5025F}). As
for the two radio halos, the Perseus radio morphology resembles that in the
X-rays. We proceed as before, but now adopt a higher central magnetic field
strength of $B_{0}=10$~$\mu$G. Such a larger $B_0$ is expected in a CCC with its
higher central gas density, implying a larger adiabatic compression factor of
the magnetic field during the condensation of the cool core (see the
\citealp{2010ApJ...710..634A,2011arXiv1111.5544M}, for a discussion on the
Perseus magnetic field).

Our parameter space study of $\gamma_{\rmn{tu}}$ and $\alpha_{\rmn{B}}$ favors
low $\gamma_{\rmn{tu}}$ values---in accordance with our expectation for
mini-halos. However, a large region of that parameter space, up to
$\gamma_{\rmn{tu}} = 100$, can equally well fit the data. The coloring of the
goodness of fit (reduced $\chi^2$) in the $\gamma_{\rmn{tu}}-\alpha_{\rmn{B}}$
plane shows the anti-correlation of $\gamma_{\rmn{tu}}$ and $\alpha_{\rmn{B}}$:
large $\gamma_{\rmn{tu}}$ values (peaked CR profiles) and low $\alpha_{\rmn{B}}$
values (flat magnetic profiles) combine to match the observed surface brightness
profile and vice versa.

In Fig.~\ref{fig:SBmodeling}, we show the two parameter combinations
($\gamma_{\rmn{tu}}=3$, $\alpha_{\rmn{B}}=0.4$) and ($\gamma_{\rmn{tu}}=100$,
$\alpha_{\rmn{B}}=0.3$). Both model realizations nicely recover the surface
brightness profile and the total luminosity within 10 per cent.  The gamma-ray
flux within $R_{200}$ for the $\gamma_{\rmn{tu}}=3$ case and for energies above
100~MeV (100~GeV) is $F_{\gamma} = 1.4 \times 10^{-8}$ ($5.1 \times 10^{-12}$)
cm$^{-2}$~s$^{-1}$. Adopting $\gamma_{\rmn{tu}}=100$ and $
\alpha_{\rmn{B}}=0.3$, the corresponding gamma-ray flux above 100~MeV (100~GeV)
is $F_{\gamma} = 4.9 \times 10^{-9}$ ($1.8 \times 10^{-12}$)
cm$^{-2}$~s$^{-1}$. Note that \emph{Fermi}-LAT measured the gamma-ray flux above
100~MeV of the central galaxy NGC~1275 to $2 \times 10^{-7}$~cm$^{-2}$~s$^{-1}$
\citep{2009arXiv0904.1904T}, well above our model predictions due to
hadronically produced diffuse gamma-ray emission that is expected to mostly glow
from the core region of the cluster.

We can compare these predictions with the upper limit above 1~TeV, and for a
region within $0.15$~deg around the cluster center, recently obtained by the
\cite{2011arXiv1111.5544M}. For $\gamma_{\rmn{tu}}=3$ ($\gamma_{\rmn{tu}}=100$),
we obtain a flux of $F_{\gamma}(>1~\rmn{TeV},<0.15~\rmn{deg}) = 7.3 \times
10^{-14}$ ($5.5 \times 10^{-14}$) cm$^{-2}$~s$^{-1}$, which is well below the
upper limit of the MAGIC collaboration,
$F_{\gamma,\rmn{UL}}(>1~\rmn{TeV},<0.15~\rmn{deg}) \approx 1.4 \times
10^{-13}$~cm$^{-2}$~s$^{-1}$. Note also that, in the case of $\gamma_{\rmn{tu}}
= 100$, we obtain a maximum CR acceleration efficiency multiplier of
$g(\zeta_{\rmn{p,max}})=0.52$, about half of the value adopted by
\cite{2010MNRAS.409..449P}. Note that adopting $g(\zeta_{\rmn{p,max}})=1$
results in slightly smaller gamma-ray luminosities in comparison to those
predicted by \cite{2010MNRAS.409..449P} and \cite{2011arXiv1105.3240P} because
we additionally account for the central temperature dip and as well as the
decrease towards larger radii.

\subsection{The Ophiuchus radio mini-halo}

The Ophiuchus cluster has been widely studied both in radio and X-rays in the
last few years because of the claimed presence of a non-thermal hard X-ray tail
\citep{2008A&A...479...27E,2008PASJ...60.1133F,2009A&A...499..371G,
  2009A&A...499..679M,2009MNRAS.396.2237P,2009A&A...508.1161N,2010A&A...514A..76M,
  2010MNRAS.405.1624M}.  It was classified as a merging cluster by
\cite{2001PASJ...53..605W}, but more recently \cite{2008PASJ...60.1133F} did not
find any evidence of merging and, on the contrary, classified it as one of the
hottest clusters with a cool-core (see also \citealp{2010MNRAS.405.1624M}).  To
simplify modeling, we neglect the small central temperature dip for radii
$r<30\,h_{70}^{-1}$~kpc and adopt a constant central temperature. (Owing to its
small size, the cool core region has no influence on the resulting radio surface
brightness.) Again, the radio mini-halo morphology displays similarities with
the thermal X-ray emission. For our modeling, we use the surface brightness
profile provided by \cite{2009A&A...499..679M}.

We proceed as before, adopting a central magnetic field value of
$B_{0}=10$~$\mu$G. Similarly to Perseus, low $\gamma_{\rmn{tu}}$ values are
favored, as expected for mini-halos. However, large regions of the parameter
space provide excellent fits to the data.  In Fig.~\ref{fig:SBmodeling}, we show the
two parameter combinations ($\gamma_{\rmn{tu}}=3$, $ \alpha_{\rmn{B}}=0.4$) and
($\gamma_{\rmn{tu}}=100$, $ \alpha_{\rmn{B}}=0.3$).  For those, we recover the surface
brightness profile and the total luminosity within 20 per cent. The gamma-ray flux
within $R_{200}$ for the $\gamma_{\rmn{tu}}=3$ case and for energies above
100~MeV (100~GeV) is $F_{\gamma} = 1.3 \times 10^{-10}$ ($4.9 \times 10^{-14}$)
cm$^{-2}$~s$^{-1}$. Adopting $\gamma_{\rmn{tu}}=100$ and $ \alpha_{\rmn{B}}=0.3$, the
corresponding gamma-ray flux above 100~MeV (100~GeV) is $F_{\gamma} = 8.3 \times
10^{-11}$ ($3.1 \times 10^{-14}$) cm$^{-2}$~s$^{-1}$. The gamma-ray flux is, in
both cases, about two orders of magnitude lower than the upper limit obtained by
\emph{Fermi}-LAT \citep{2010ApJ...717L..71A}. Note also that in the case of
$\gamma_{\rmn{tu}} = 100$ we obtain a maximum CR acceleration efficiency
multiplier of $g(\zeta_{\rmn{p,max}})=0.014$.


\section{Discussion: a hybrid scenario for Giant and Mini Radio Halos?}
\label{sec:discussion_hadronic}

In order to cleanly assess the possibility of the hadronic model to alone
explain the RH data, we only considered the hadronically-induced radio emission
component in the preceding section. Hence, by construction, we neglected other
(leptonic) emission components, such as reaccelerated electrons.  We now
  address possible biases that may have affected our previous conclusions.

\subsection{Biases of the hadronic model of  radio halos}

(i) Merging clusters are not spherically symmetric as can be seen in Coma and
Abell~2163, requiring inherently non-spherical modeling. In order to reproduce
the more extended radio emission relative to the thermal X-ray emission, the
non-thermal clumping factor, $C_{\rmn{non-th}}$, needs to be larger than its
thermal analogue, $C_{\rmn{th}}$, in concentric spherical shells, where we
defined those statistics by
\begin{eqnarray}
  \label{eq:clumping}
  C_{\rmn{non-th}} &=&
  \expval{\rho_\rmn{gas} C}/\expval{\sqrt{\rho_\rmn{gas} C}}^2,\\
  C_{\rmn{th}} &=& 
  \expval{\rho_\rmn{gas}^2}/\expval{\rho_\rmn{gas}}^2.
\end{eqnarray}
This manifests itself, e.g., in the large-scale morphology of the radio
surface brightness emission, which is more elongated than its counterpart in
thermal X-rays, but also on scales smaller than the radio beam. In our
phenomenological modeling, we allow for those deviations by means of the
parameters $\gamma_{\rmn{tu}}$ and $ \alpha_{\rmn{B}}$ for the CRs and
magnetic fields, respectively.  While this approach is well suited to describe
large-scale anisotropies, it may be inadequate to model small-scale
inhomogeneities such as CR trapping in magnetic mirrors through the second
adiabatic invariant and needs to be carefully quantified in future work.

(ii) Adopting the simulation-derived $\tilde{C}$ profile
\citep{2010MNRAS.409..449P} for our extended model may have biased the inner
slope of the CR density profile to become too steep due to the overcooling
problem of purely radiative simulations. This produces cluster cores that are
too dense (in comparison to observations), which also should overestimate the
rate of adiabatic compression that is experienced by the CR population during
the formation of the cooling core. Hence, the resulting values of
$\gamma_{\rmn{tu}}$ are then biased low in comparison to a potentially shallower
slope of the inner CR profile. To quantify the last point, we try to reproduce
the Coma surface brightness at 1.4 GHz using a model without $\tilde{C}$. We
find that values as high as $\gamma_{\rmn{tu}} \approx 8$ can be
accommodated. However, $\gamma_{\rmn{tu}}=1$ still represents the best match to
the data, demonstrating that the problem can be weakened but not circumvented
even in this case of a cored CR profile.

(iii) Considering the case of advection-dominated CR transport
($\gamma_{\rmn{tu}}\gtrsim100$), which only allows for a good match to the
mini-halo data of Perseus and Ophiuchus, the $g_{\rmn{CR}}$ parameter can be
interpreted as the maximum CR acceleration efficiency used in
\cite{2010MNRAS.409..449P}. If the cluster CR population is mainly accelerated
in cosmological structure formation shocks, then this value should depend on the
mass accretion history and should be approximately universal, i.e., similar for
all clusters. We find $g_{\rmn{CR, Perseus}} = 0.52$ and $g_{\rmn{CR,
    Ophiuchus}} = 0.014$, because we fixed $B_{0}=10$~$\mu$G in both cases and
used $g_{\rmn{CR}}$ as normalization. This discrepancy can be resolved by
increasing/lowering the central magnetic field in Perseus/Ophiuchus to
$B_{0,\rmn{Perseus}}\approx20$~$\mu$G and
$B_{0,\rmn{Ophiuchus}}\approx1$~$\mu$G. We note, however, that without the
guidance of cosmological cluster simulations that include CR transport, the data
does not yet constrain $\gamma_{\rmn{tu}}$.

The small cluster sample analyzed here is only meant to serve as a proof of
concept and to show the viability of matching observed representative RH data
with our extended hadronic model. However, it seems unlikely that the biases
discussed above severely affect our findings that the extended hadronic model
successfully reproduces the main morphological characteristics of radio
mini-halos with a wide range of possible values for $\gamma_{\rmn{tu}}$ and
without violating gamma-ray constraints. In contrast, the hadronic model appears
to fail in explaining the radio emission in the outskirts of the Coma RH at low
frequencies and requires a flat CR distribution both in Coma and in A2163 at
1.4~GHz. This motivates us to consider a modification of this purely
  hadronic model in explaining RHs.

\subsection{Hybrid hadronic-leptonic model}

Within the hadronic scenario, there emerges a plausible physical solution
  to this observational challenge. We suggest that the rich phenomenology of RHs
  may be a consequence of two different radio emission components---one of which
  is induced by hadronic interactions and the other is of leptonic origin
  \citep{2008MNRAS.385.1211P}.  There are a number of plausible processes for
  the latter.  These includes turbulent re-acceleration of primary or secondary
  (hadronically produced) electrons \citep{2010arXiv1008.0184B} or
  re-acceleration of fossil electrons by means of diffusive shock acceleration
  \citep{kang11,kang12,pinzke13}. The fossil electron population may originate
  from the time-integrated and successively cooled population of directly
  injected electrons at strong structure formation shocks that the gas
  experienced trough it cosmic accretion history. Alternatively, a seed
  population of relativistic electrons could be provided by the time-integrated
  action of AGN feedback or by supernova-driven galactic winds.  Depending on
  relative strength of the different components, this scenario would imply
  various halo phenomena:
\begin{enumerate}
\item A dominating hadronic component  manifests in form of radio mini-halos in
  CCCs \citep{2004A&A...413...17P}.
\item When the leptonic component dominates, we should have steep spectrum halo
  sources \citep[such as A520,][]{2008Natur.455..944B}, some of which could be
  produced by giant radio relic sources projected onto the main cluster
  \citep{2012arXiv1211.3122S}.
\item The case of both components significantly contributing to the diffuse
  radio emission results in giant radio halos, with the hadronic component
  dominating in the center and the leptonic emission taking over in the outer
  parts. The peripheral regions of merging clusters experience an especially
  high level of kinetic pressure contribution \citep{2009ApJ...705.1129L,
    2012ApJ...758...74B} that manifests in form of subsonic turbulence (as
  suggested observationally by \citealp{2004A&A...426..387S} or theoretically by
  \citealp{2006MNRAS.366.1437S,2005MNRAS.364..753D, 2008Sci...320..909R}) and a
  complex network of shocks \citep{2003ApJ...593..599R, 2006MNRAS.367..113P,
    2008MNRAS.385.1211P, 2008ApJ...689.1063S, 2009MNRAS.395.1333V}. Depending on
  the merger geometry and dynamical stage, as well as on the electron
  acceleration efficiencies of these non-equilibrium processes and the CR
  streaming speeds, we would expect the development of a (fuzzy) transition
  region between hadronic and leptonic component. This generalizes the
  simulation-inspired model by \cite{2008MNRAS.385.1211P} who propose that
  primary electron substantially contribute to the peripheral RH emission.
\end{enumerate}

  A detailed implementation of this hybrid scenario would depend very much on
  the precise characterization of a given cluster. This goes beyond the scope of
  the present work, which mainly explores the possible observational
  consequences of the hadronic component for future radio surveys that are
  implied by our extended model. Nevertheless, we sketch possible observational
  implications of a hybrid hadronic-leptonic scenario in the following
  subsection.

\subsection{Observational implications}

\subsubsection{Spectral and morphological variability}
In mini-halos and in the centers of giant halos, where the hadronic component
dominates in our picture, we would naively expect at most modest spectral
variations. This is because these regions average over sufficiently many fluid
elements, each of which experienced its characteristic shock history during the
cluster assembly. However, when averaged over the ensemble, this produces a CR
population that has a nearly universal spectrum
\citep{2010MNRAS.409..449P}. However, CR streaming and diffusive transport may
cause a possible spectral steepening in the cluster core region because more
energetic CRs diffuse/stream faster. This would then imply spatial variations of
the CR spectral index and, hence, spatially varying radio emission throughout
the cluster (core region) when taking the CR advection effects into account,
which would mix regions of different CR spectral properties. In regions where
the leptonic component dominates (such as the outer regions of giant halos or
steep spectrum halo sources), we expect substantial spectral and morphological
variations in the radio maps. This is because of the intermittency of the
acceleration process (acceleration at discrete weak shocks or turbulent
acceleration), the expected distribution of Mach numbers or CR momentum
diffusion coefficient, respectively, and the comparably short electron cooling
time ($\sim100$~Myr). Interestingly, this compares well with the large azimuthal
scatter of different sector profiles of the Coma halo, Fig.~4 of
\cite{2011MNRAS.412....2B}, and fronts (primarily towards the West) in their
high-resolution surface brightness map, which may indicate the transition from
the hadronic to the leptonic emission component. In particular, the relative
inefficiency of shock acceleration at weak shocks or turbulent acceleration
generates steeper radio spectra in the leptonically dominated regions. Hence,
this would naturally imply a radial spectral steepening and cause substantial
morphological and spectral variation in the outer regions of giant halos. The
increasing fraction of the leptonic component towards lower frequencies may then
imply a larger halo size with decreasing observational frequency.

\subsubsection{Halo switch-on/-off mechanism and the radio-X-ray bimodality}
Clearly, a cluster merger injects turbulence and shocks that could both
accelerate fossil electrons and switch the leptonic component on. On the other
hand, CR advection produces centrally enhanced CR profiles because of adiabatic
compression of CRs for radial eddies. The energetization and transport of CRs to
the central halo regions implies a lightening up of the hadronic emission
component. For the leptonic component, the halo switch off is faster or
comparable to the dynamical time scale, $t_{\rm dyn} \sim t_{\rm
  H}/\sqrt{\Delta} \sim 1 \,\rmn{Gyr} \Delta_{100}^{-1/2}$, where $t_{\rm
  H}=10$~Gyr, $\Delta_{100}=\rho/(100\bar{\rho})$. In the case of diffusive
shock acceleration of fossil electrons, the radio emission will be shut off
within a CR electron cooling time ($t_\rmn{cool}\sim100$~Myr) if the
acceleration source ceases, i.e., when shocks have dissipated all the energy. In
the case of the turbulent re-acceleration model, the turbulence decays on a few
eddy turnover time scales on the injection scale which should take somewhat
longer. The hadronic emission component is also expected to decrease
substantially once turbulent pumping of CRs ceases and CRs are set free to
stream, which results in a net CR flux towards the external cluster regions. The
accompanying flattening of the CR profile implies a lowering of the hadronic
radio emission because the CRs see a smaller target density in the outer parts.
This should lead naturally to a bimodality of radio synchrotron emissivities due
to hadronic and leptonic halo components.


\begin{figure*} 
\centering
\includegraphics[width=0.49\textwidth]{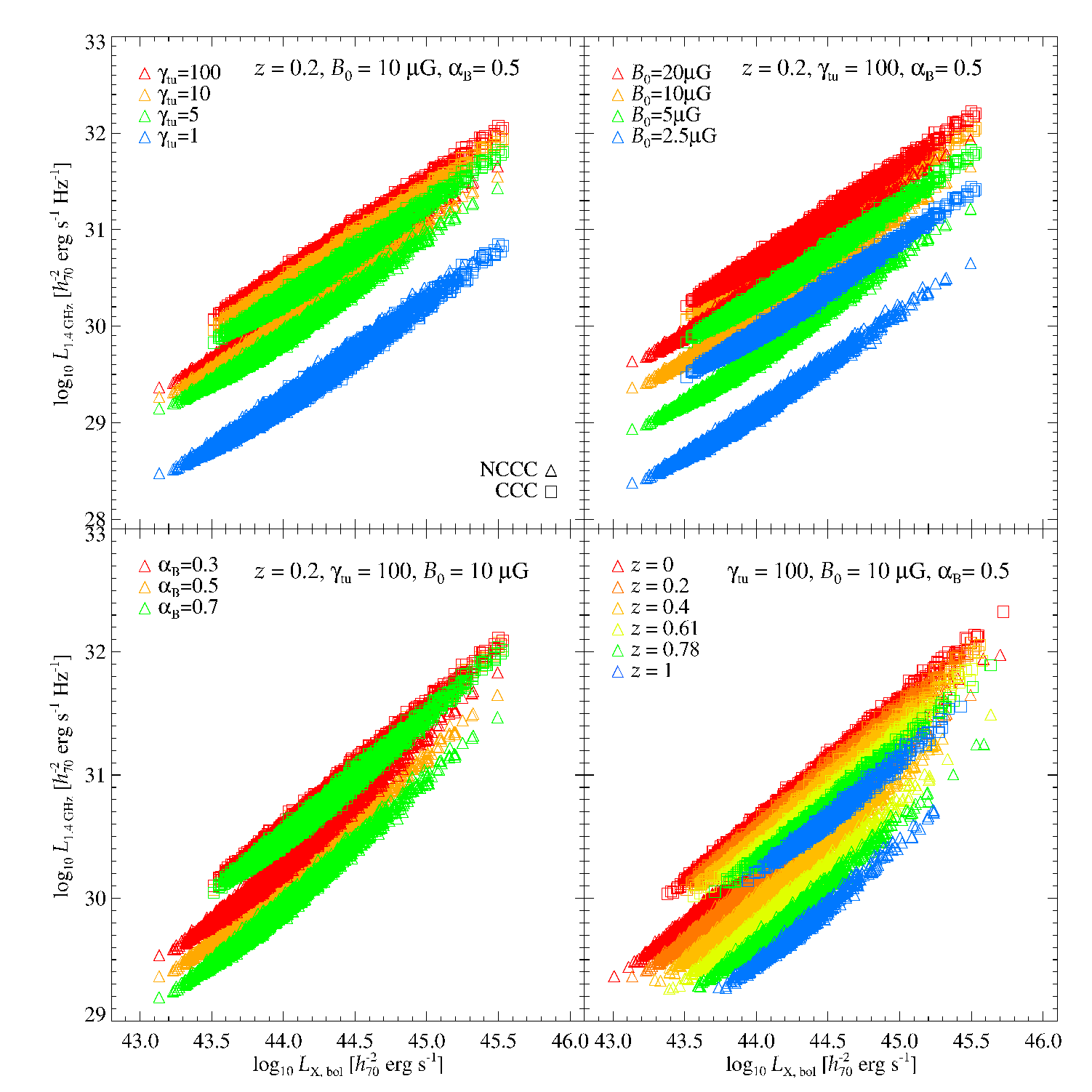}
\includegraphics[width=0.49\textwidth]{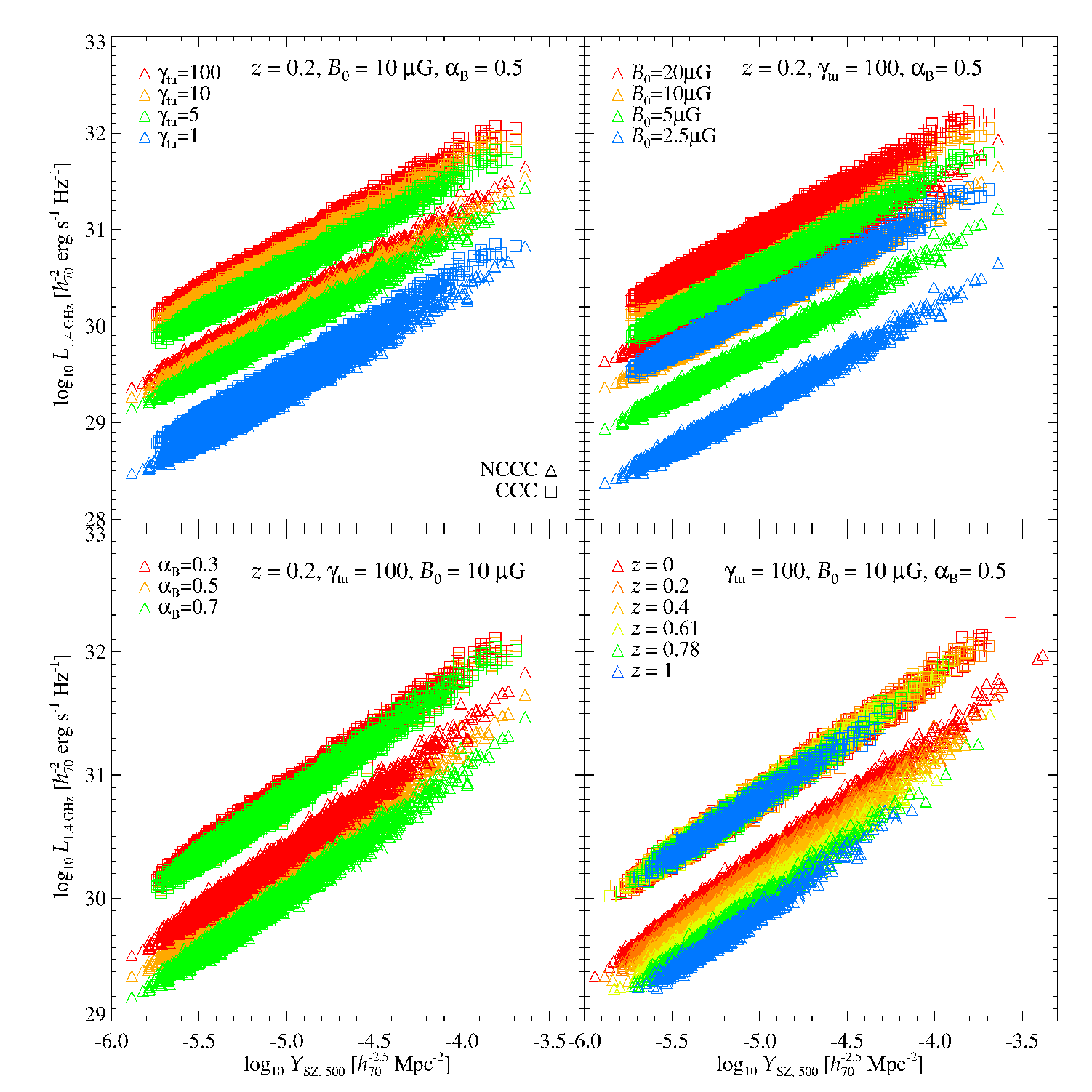}
\caption{Radio-to-X-ray and radio-to-SZ scaling relations as predicted by our
  extended CR model. In the left panel, we show how the
  $L_{1.4~\rmn{GHz}}-L_{\rmn{X,bol}}$ relation varies upon changing different
  parameters. In the right panel, we show the same, but for the
  $L_{1.4~\rmn{GHz}}-Y_{\rmn{SZ},500}$ relation. Note that in each plot there
  are two separated populations for each model realization, shown with the same
  color but different symbols. The upper sets of points (squares) correspond to
  the CCC population while the lower sets (triangles) correspond to NCCCs. The
  plot labels indicate those parameters which are kept fixed. We also fix the
  $g_{\rmn{CR}}$-normalization parameter to 0.5 for all cases. See main text for
  further details.}
\label{fig:SR}
\end{figure*}

\section{Scaling Relations}
\label{sec:4}
As introduced in Section~\ref{sec:1}, there exist an apparent bimodality between
the radio and X-ray cluster emission. Clusters with a given X-ray luminosity can
either host RHs or show an absence of diffuse radio emission (e.g.,
\citealp{2009A&A...507..661B,2011A&A...527A..99E}). More recently, a study of
the radio-to-SZ scaling relation revealed the absence of a strong bimodality
dividing the cluster population into radio-loud and radio-quiet clusters
\citep{2012MNRAS.421L.112B,2013arXiv1306.4379C,2013arXiv1307.3049S}.  Since
$Y_{\rmn{SZ}}$ correlates more tightly with cluster mass than $L_{\rmn{X}}$,
this may indicate that the larger scatter of $L_{\rmn{X}}$ correlates with the
scatter of the radio luminosity in such a way that it produces a bimodality; but
as a result of a second (hidden parameter) rather than the cluster mass. In this
section, we investigate these two scaling relations in the framework of our
extended hadronic scenario.

  In the following, we apply our model to the complete cosmological mock
  cluster catalog build from the MultiDark $N$-body simulation in our
  Paper~I. For each object in the sample, we use the cluster mass, a dynamical
  disturbance parameter (the normalized distance of the halo center and the
  center of mass) for sorting the cluster into the CCC/NCCC populations, and a
  phenomenologically assigned ICM density to calculate the radio (and gamma-ray)
  emission.

\subsection{Exploring the parameter space of scaling relations}

In Fig.~\ref{fig:SR}, we show the general scaling relations of our extended CR
model of Section~\ref{sec:2.3} applied to the MultiDark sample. We show how both
the radio-to-X-ray and the radio-to-SZ scaling relations differ upon varying the
parameters $\gamma_{\rmn{tu}}$, $B_{0}$, $\alpha_{\rmn{B}}$ and redshift. We fix
the CR-normalization parameter $g_{\rmn{CR}}$ to 0.5 in all cases, ensuring an
average CR-to-thermal pressure of 2 per cent (0.05 per cent) within $R_{500}$
($R_{500}/2$). Here, the radio luminosity is calculated at $1.4$~GHz within
$R_{500}.$\footnote{The mean (median) difference between calculating $L_{\nu}$
  within $R_{200}$ or $R_{500}$ is 5.3 per cent (5.6 per cent).}  In our CR
model, we fix the CR number for $\gamma_{\rmn{tu}}=100$ using equation~(36) of
\cite{2011A&A...527A..99E}, integrating up to $R_{500}$. To compute the radio
luminosity for different values of $\gamma_{\rmn{tu}}$, we employ CR number
conservation (for CR energies $E>$~GeV where Coulomb cooling is negligible for
CR protons).

In each panel in Fig.~\ref{fig:SR} there are two separated populations for
each model realization (i.e., for a given set of parameters). Each upper set of
points (squares) corresponds to the CCC population while the lower set (triangles) 
corresponds to NCCCs, respectively. In our model, the radio and X-ray emissivities scale 
with the square of the gas density so that $L_{1.4~\rmn{GHz}}$ and $L_{\rmn{X,bol}}$
are significantly higher for CCCs in comparison to NCCCs. In contrast,
$Y_{\rmn{SZ}}$ only depends weakly on the central gas density as discussed in
Paper~I. This explains the relative location of the NCCC and
CCC populations in the $L_{1.4~\rmn{GHz}}-L_{\rmn{X,bol}}$ and
$L_{1.4~\rmn{GHz}}-Y_{\rmn{SZ}}$ planes. In particular, CCCs are shifted to the
upper right in the $L_{1.4~\rmn{GHz}}-L_{\rmn{X,bol}}$ plane while they are
shifted vertically upward in the plane spanned by
$L_{1.4~\rmn{GHz}}-Y_{\rmn{SZ}}$. In reality, we expect an (ab initio unknown)
distribution of these parameters which would substantially increase the scatter
in the scaling relations and possibly lead to a bimodality, depending on
correlations among the different parameters.

Most interestingly, the slope of the radio scaling relations does not differ
when varying parameter values because we do not include any cluster
mass-dependence in our parametrizations which is not constrained by current
data. Closely inspecting Fig.~\ref{fig:SR}, we see that we obtain the largest
changes in $L_{1.4~\rmn{GHz}}$ for variations in $1<\gamma_{\rmn{tu}}<5$ and
$B_0$ over the parameter range probed, albeit with a stronger dependence for
weaker field strengths (as expected from the $B^2/(B^2+B_\rmn{CMB}^2)$ term of
equation~(\ref{eq:jnu}), where $B_\rmn{CMB}\simeq 3.2\,\umu\rmn{G} (1+z)^2$ is
the equivalent magnetic field strength of the cosmic microwave background).

\subsection{Comparison to observations}
\label{sec:scaling-obs}
 
After collecting the X-ray luminosity and the SZ flux of known RHs, we compare
the resulting scaling relations to a phenomenological model realization that was
chosen to additionally obey other observational constraints (e.g., from Faraday
rotation measure studies) as well as theoretical considerations on CR transport.

\subsubsection{Observational samples}

\begin{figure*} 
\centering
\includegraphics[width=0.49\textwidth]{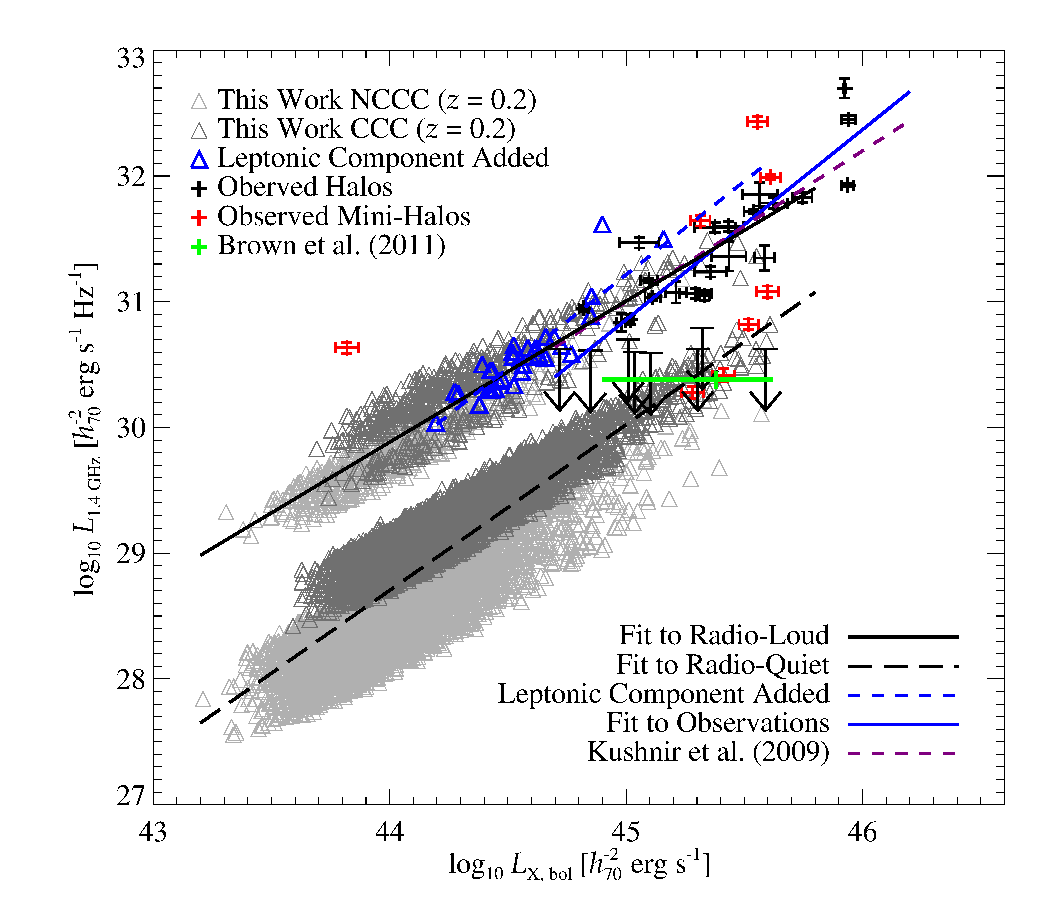}
\includegraphics[width=0.49\textwidth]{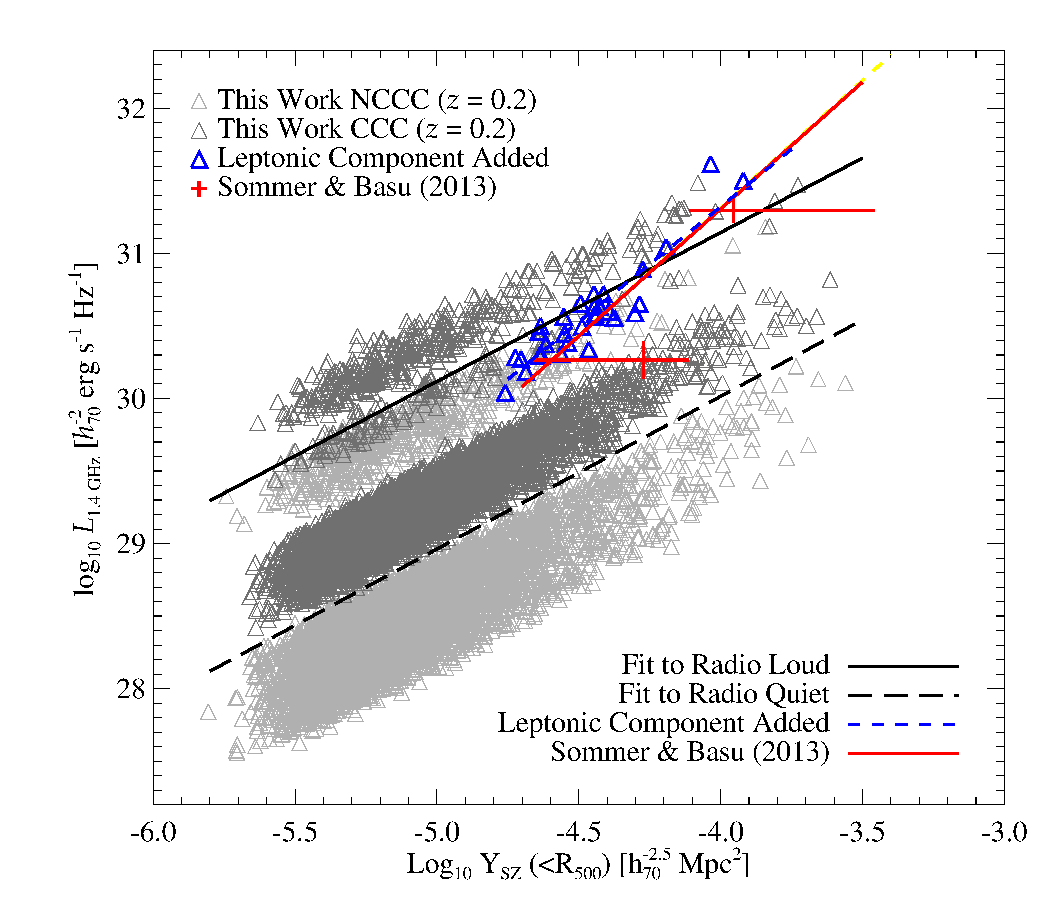}
\caption{Radio-to-X-ray and radio-to-SZ scaling relations in our extended CR model
  (see main text for the details of the chosen parameters) compared with
  observations.  \emph{Left.} $L_{1.4~\rmn{GHz}}-L_{\rmn{X,bol}}$ relation in
  comparison to the observational sample taken from the literature and detailed
  in Appendix~\ref{app:D}. We additionally show the detected signal of Mpc-scale 
  diffuse emission in a stacked sample of radio-quiet galaxy clusters 
  (green, \protect\citealp{2011ApJ...740L..28B}). 
  \emph{Right.} $L_{1.4~\rmn{GHz}}-Y_{\rmn{SZ}}$ relation 
  in comparison with
  the PSZ(V) sample (direct integration method) 
  of \protect\cite{2013arXiv1307.3049S}
  together with the data points shown in their Fig.~13.}
\label{fig:PLSZ}
\end{figure*}

In Appendix~\ref{app:D} we construct a sample of giant radio halos (black) and
radio mini-halos (red), as well as upper limits on the radio emission
\citep{2009A&A...499..371G,2009A&A...507..661B, 2011A&A...527A..99E}, and show
this in the left panel of Fig.~\ref{fig:PLSZ}. The median redshift of this
sample is $z\approx0.18$. The corresponding observational scaling relation is
well fit by $\log_{10} L_{1.4~\rmn{GHz}} = A + B~\log_{10} L_{\rmn{X,bol}}$ with
$A=-37.204\pm1.838$ and $B=1.512\pm0.041$, and a scatter of $\sigma_{yx} \approx
0.52$ (we do not include upper limits in the fit; units are as in
Fig.~\ref{fig:PLSZ}). We refer the reader to \cite{2009A&A...507..661B}, 
\cite{2011A&A...527A..99E} and \cite{2013arXiv1306.4379C} for an extensive discussion on this topic. 
We emphasize that in contrast to giant radio halos, mini-halos span a wider range
in radio luminosity (as also pointed out by \citealp{2009A&A...499..679M}). The
Perseus mini-halo (highest radio mini-halo luminosity in the left panel of
Fig.~\ref{fig:PLSZ}), e.g., has a radio luminosity that is almost an order of
magnitude higher than in giant radio halos at the same X-ray luminosity. In
contrast, the Ophiuchus mini-halo (lowest radio mini-halo luminosity in the left
panel of Fig.~\ref{fig:PLSZ}), which is representative of a few other similar
examples recently detected in CCCs (such as A2029 and A1835), has a radio
luminosity which is much lower than giant radio halos in merging clusters and is
even below the upper limits.

We caution that the determination of the slope of the observational
$L_{1.4}-L_{\rmn{X,bol}}$ relation is not very robust because of the small
sample size of RHs, selection biases of extended low-surface brightness objects,
and systematic uncertainties in the measurements of $L_{1.4}$ and
$L_{\rmn{X,bol}}$. The fact that there have been low-luminosity mini-halos found
only recently \citep{Giacintucci} exemplifies those uncertainties and the large
intrinsic scatter of this relation. On the other hand, X-ray luminosities for
the same object as derived by, e.g.,~\emph{ROSAT} and \emph{Chandra} can easily
differ by a factor of up to a few.\footnote{For example, the bolometric X-ray
  luminosity of A2163 as measured by \emph{ROSAT} is
  $8.65\times10^{45}$~$h_{70}^{-1}$~erg~s$^{-1}$ \citep{2009A&A...507..661B}
  while the \emph{Chandra} measurement is
  $4.93\times10^{45}$~$h_{70}^{-1}$~erg~s$^{-1}$ (\citealp{2009ApJS..182...12C};
  ACCEPT: Archive of Chandra Cluster Entropy Profile Tables;
  http://www.pa.msu.edu/astro/MC2/accept/).}  In the left panel of
Fig.~\ref{fig:PLSZ}, we additionally show the model of
\citet{2009JCAP...09..024K} with a slope of $\approx1.2$, arbitrarily normalized
for visual purposes, from their simple analytical hadronic model.

  In order to compare our model to the observed 1.4~GHz radio-to-SZ scaling
  relation, we use the result by \cite{2013arXiv1307.3049S} that is based on a
  sub-sample of the \emph{Planck} COSMO sample \citep{2013arXiv1303.5080P} with
  a median redshift of $z \approx 0.22$ (we use their PSZ(V) sample), which
  compares favorably with our MultiDark $z = 0.2$ snapshot. The same comments
  regarding the small sample size of RHs, selection biases, and systematic
  uncertainties in the luminosity measurements also apply here.

\subsubsection{Model realization}

In order to compare with observations, we select a particular realization of our
extended CR model. To this end, we use the MultiDark cluster sample at $z=0.2$,
which compares well with the redshift of the observational samples (see above
and Appendix~\ref{app:D}). We divide our cluster sample randomly into
radio-quiet and radio-loud clusters, assuming a ratio of 10 per cent of the
latter (see next section). In our model, we use the turbulent propagation
parameter $\gamma_{\rmn{tu}}$ to separate both populations. In radio-quiet
clusters, we assign $\gamma_{\rmn{tu}}=1$, and in radio-loud clusters, we adopt
randomly and uniformly $\gamma_{\rmn{tu}}$ values in the intervals $[40,80]$ and
$[1,5]$ for NCCCs and CCCs, respectively.

Our modeling of magnetic fields is inspired by Faraday rotation studies that
point to higher field values in the core region of CCCs compared to NCCCs
\citep{2010A&A...513A..30B,2011A&A...529A..13K}, presumably due to the higher
adiabatic compression factor during the formation of the cooling core. Hence,
for radio-quiet clusters, we adopt randomly and uniformly distributed values of
the central magnetic field $B_0$ in the intervals $[2.5,5.5]$~$\umu$G and
$[5,10]$~$\umu$G for NCCCs and CCCs, respectively. To account for the potential
turbulent dynamo in radio-loud objects (characterized by a higher turbulent
transport parameter in our model), we slightly increase $B_0$ in those objects
and chose $B_0$ intervals of $[4.5,7.5]$~$\umu$G and $[7.5,12.5]$~$\umu$G for
NCCCs and CCCs, respectively.

We fix $\alpha_{\rmn{B}}=0.5$ and $g_{\rmn{CR}}=0.5$ for all clusters. We note
that our parameter choices are mostly phenomenologically driven with the aim to
reproduce observations. The parameter study in Fig.~\ref{fig:SR} exemplifies
considerable degeneracies so that different combinations of parameters can
potentially result in very similar distributions. We emphasize the need of more
detailed observations of RHs and in particular of multi-frequency
correlation studies to constrain the interplay of some of these parameters.

In Fig.~\ref{fig:PLSZ}, we show our model in comparison to the observed
radio-to-X-ray and radio-to-SZ scaling relations. The normalization of our model
can be arbitrarily varied by changing $g_{\rmn{CR}}$ as long as the resulting
$X_{\rmn{CR}}$ respects the current observational constraints and remains below
a few percents. As explained above, our choice of $g_{CR}=0.5$ ensures an
average CR-to-thermal pressure of 2 per cent within $R_{500}$.

Our model is sufficiently flexible to either mimic a cluster radio bimodality or
not, depending on the parameters adopted for the populations of radio-loud and
radio-quiet objects. However, with the given model realization as in
Fig.~\ref{fig:PLSZ}, the separation of the radio-loud and radio-quiet
populations is substantially larger in the $L_{1.4~\rmn{GHz}}-L_{\rmn{X,bol}}$
plane than in the $L_{1.4~\rmn{GHz}}-Y_{\rmn{SZ}}$ plane, which exhibits almost
a continuum distribution from radio-loud CCCs to the radio-quiet NCCCs. This is
mainly because the bolometric X-ray emissivity scales with $\rho_{\rmn{gas}}^2$
while $Y_{\rmn{SZ}}\propto\rho_{\rmn{gas}}$ (which is only strictly valid for an
isothermal gas distribution).  This is one plausible explanation for the
observed discrepancy of the presence of a bimodality in
$L_{1.4~\rmn{GHz}}-L_{\rmn{X,bol}}$ and the apparent absence of it in
$L_{1.4~\rmn{GHz}}-Y_{\rmn{SZ}}$.

The slope of our model depends on the different parameter choices and,
particularly, on the relative differences introduced for the NCCC/CCC and the
radio-loud/quiet populations. However, we note that our
$L_{1.4~\rmn{GHz}}-L_{\rmn{X,bol}}$ and $L_{1.4~\rmn{GHz}}-Y_{\rmn{SZ}}$ scaling
relations are somewhat shallower than the observed relation, more similar to the
model by \cite{2009JCAP...09..024K}.  This may hint at the contribution of
  a second, leptonic component that would steepen the slope of our model scaling
  relation at high mass.  In particular, the finding of
  \cite{2013arXiv1306.4379C} that clusters do seem to show a bimodality at very
  high $Y_{\rmn{SZ}}$ may also be an hint of an increasingly important leptonic
  component. However, we do not expect this component to significantly alter our
  conclusions regarding the luminosity functions of the next section for the
  following reasons. (i) The leptonic component would only be present in the
  radio-loud merging NCCC sample, i.e., the radio (giant) halos, and (ii) it
  would only be dominant at very high masses because the dissipated energy that
  is available for energizing fossil electrons should be a fraction of the
  thermal energy which scales with cluster mass as $E_\rmn{th}\propto
  M_{200}^{5/3}$ \citep[][for magneto-turbulent re-acceleration
  models]{2005MNRAS.357.1313C}. In fact, our $z = 0.2$ mock sample, only
  contains $31$ radio-loud \mbox{NCCCs} with
  $M_{200}\geq5\times10^{14}~h_{70}^{-1}$~M$_{\odot}$, which is the mass range
  of, e.g., turbulently reaccelerated RHs \citep{2010A&A...509A..68C}.

  To visualize such a possible leptonic component, we boost the total flux of
  these 31 radio-loud NCCCs according to $L_{1.4~\rmn{GHz, boosted}} =
  L_{1.4~\rmn{GHz, had}} + L_{1.4~\rmn{GHz, boost}}$, where $L_{1.4~\rmn{GHz,
      had}}$ is the radio luminosity of our extended hadronic component and
\begin{equation}
L_{1.4~\rmn{GHz, boost}} = L_{1.4~\rmn{GHz, had}} \times 
\left( \frac{M_{500}}{7.5\times10^{14}~h_{70}^{-1}~\rmn{M}_{\odot}} \right)^{2.3} \propto M_{500}^4 \, .
\label{eg:leptonic_boost}
\end{equation}
We consider this to be a phenomenological correction factor that aims at
reproducing the observed relation $L_{1.4~\rmn{GHz}} \propto M_{500}^4$
\citep{2007MNRAS.378.1565C,2013arXiv1306.4379C}. Possible physical realizations
include turbulent re-acceleration of primary or secondary (hadronically
produced) electrons \citep{2010arXiv1008.0184B} or re-acceleration of fossil
electrons by means of diffusive shock acceleration
\citep{kang11,kang12,pinzke13}. Here we tie the leptonic component to our
modeling of the magnetic field and the hadronic emission component, which
provides guidance for the missing signal fraction that we require by our surface
brightness modeling. As $L_{1.4~\rmn{GHz, had}} \propto M_{500}^{1.7}$, we adopt
an additional mass scaling to reach the desired $L_{1.4~\rmn{GHz}} \propto
M_{500}^4$.  The resulting median (mean) boost is about 32 per cent (47 per
cent) of the hadronic component.\footnote{The modeling of Coma and
    Abell~2163 suggests that a boost of about 35 per cent to 80 per cent may be
    necessary. However, if we allowed for flatter CR profiles in turbulent,
    merging clusters as in \cite{2013arXiv1303.4746W}, the required fraction of
    leptonic component could be smaller.}  The boosted population is shown in
both panels of Fig.~\ref{fig:PLSZ}. The corresponding scaling relations have
slopes close to the ones of the $L_{1.4~\rmn{GHz}}-L_{\rmn{X,bol}}$ and
$L_{1.4~\rmn{GHz}}-Y_{\rmn{SZ}}$ observational samples, respectively.

   At low SZ fluxes, there are a considerable number of radio loud radio mini
  halos visible in CCCs that fall above the observed
  $L_{1.4~\rmn{GHz}}-Y_{\rmn{SZ}}$ relation by
  \citet{2013arXiv1307.3049S}. However, this does not challenge our model
  because (i) in order to characterize the radio halo emission,
  \citet{2013arXiv1307.3049S} apply a low-pass filter to the radio data, which
  minimizes any flux contribution from radio mini-halos that are comparable or
  smaller than the chosen filter size and (ii) the radio mini-halo population in
  the literature suffers from incompleteness effects \citep{Giacintucci}.

Owing to the many uncertainties and lack of robustness both in the observations
and modeling at this stage, we do not attempt to fine-tune our model to the
observations. In particular, we refrain from introducing any
  mass-dependence in our free parameters at this stage.  Additionally, we do not
  include the possible leptonic emission component in the analysis of next
  section, deferring the study of its physical details and correlation to the
  hadronic component to future work.  The mock cluster sample used here is
  affected by incompleteness in the highest-mass range because of the limited
  volume of the MultiDark simulation (Paper~I). Only a small number of objects
  that lie in this mass range would be affected by such a correction, as
  discussed above and shown in Fig.~\ref{fig:PLSZ}. These are not statistically
  significant in comparison to the RH abundances that we will find in the next
  section.  Interestingly, the detected signal of Mpc-scale diffuse emission in
a stacked sample of radio-quiet galaxy clusters (shown in green in the left plot
of Fig.~\ref{fig:PLSZ}, \citealp{2011ApJ...740L..28B}) agrees with the expected
signal of our radio-quiet population.

\section{Luminosity Functions}
\label{sec:5}

\begin{figure*} 
\centering
\includegraphics[width=0.85\textwidth]{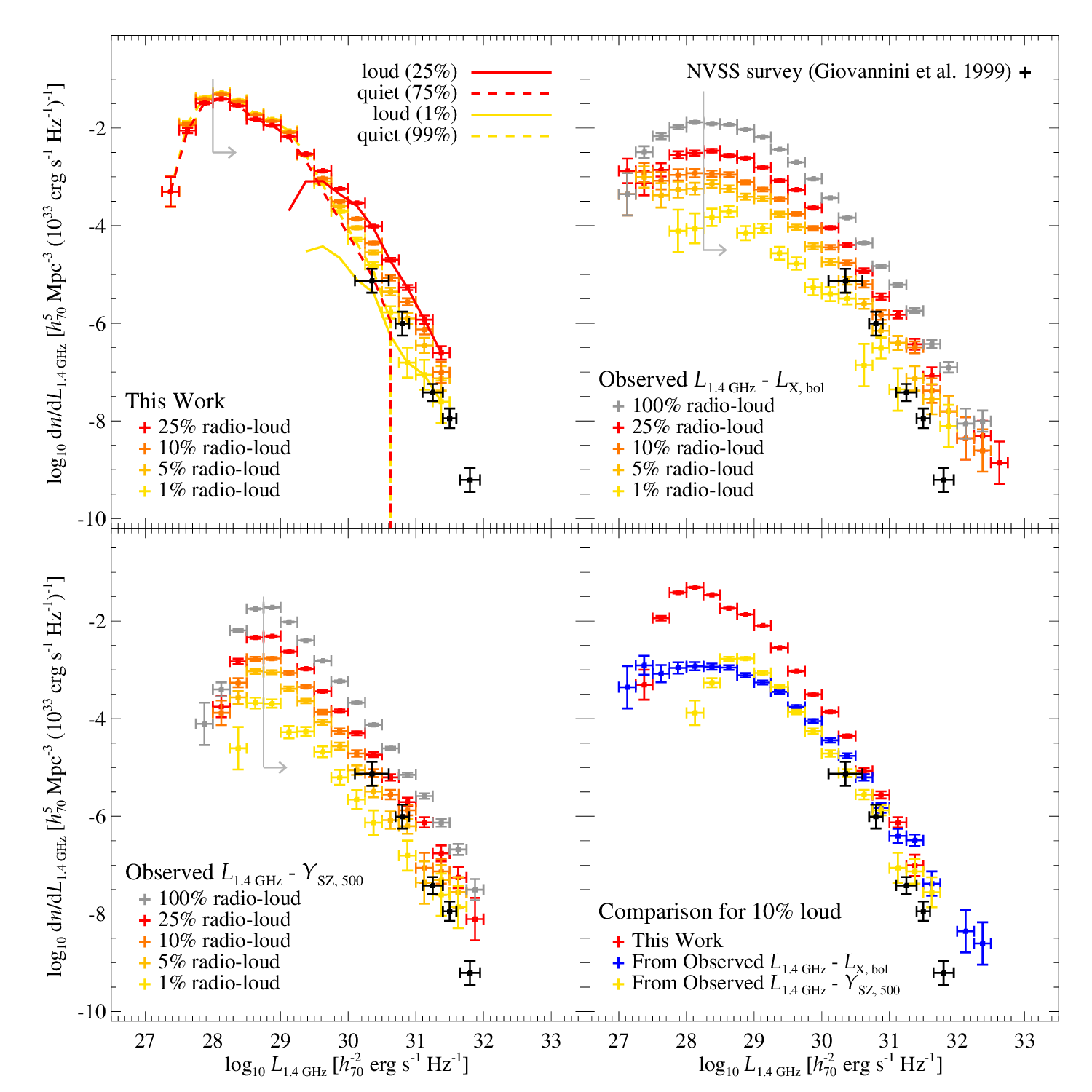}
\caption{RH luminosity function (RLF) at 1.4~GHz. The top left panel shows the
  RLF of our extended CR model (see main text for the details of the chosen
  parameters) for different fractions of radio-loud clusters. Additionally shown
  is the contribution of radio-quiet and radio-loud populations to the total RLF
  (assuming a fraction of 25 per cent and 1 per cent of radio-loud objects). The
  top right panel shows the RLF obtained by applying the observed
  $L_{1.4~\rmn{GHz}}-L_{\rmn{X,bol}}$ relation (see left panel of
    Fig.~\ref{fig:PLSZ}) to the MultiDark clusters at $z = 0.2$, using our
  phenomenological gas model for $L_{\rmn{X,bol}}$ of each cluster; again for
  different percentages of radio-loud clusters.  The bottom left panel shows the
  RLF obtained by applying the observed $L_{1.4~\rmn{GHz}}-Y_{\rmn{SZ}, 500}$
  relation (see right panel of Fig.~\ref{fig:PLSZ}) to the MultiDark
  clusters at $z = 0.2$, using $Y_{\rmn{SZ}, 500}$ of our our phenomenological
  gas model for each cluster. The bottom right panel shows the comparison
  between the three approaches for 10 per cent of radio-loud clusters. In all
  panels, we show the NVSS survey RLF \citep{1999NewA....4..141G} with a median
  redshift of $z\approx 0.18$, corrected for the sample incompleteness and
  survey sky coverage.  Horizontal error bars represent the mass bins while the
  vertical error bars are Poissonian uncertainties. The light gray line marked
  by the arrow estimates the incompleteness limit owing to the adopted low-mass
  cut (in the construction of the mock cluster catalog, see Paper~I) and the
  scatter in the halo luminosities.}
\label{fig:RLF_1.4}
\end{figure*}

\subsection{Comparison with observations at 1.4~GHz}

In Fig.~\ref{fig:RLF_1.4}, we show the RH luminosity function (RLF) at 1.4~GHz
for a representative realization of our extended CR model (as in
Section~\ref{sec:4}), and compare it with observational results.  The RLF is
completely determined by the cluster mass function and the radio
luminosity-to-mass relation, through $L_{1.4~\rmn{GHz}}-L_{ \rmn{X,bol}}$ or
$L_{1.4~\rmn{GHz}}-Y_{\rmn{SZ}}$ in combination with our phenomenological gas
model (see Paper~I). However, in the radio band there is the additional
uncertainty of the fraction of radio-loud clusters. Thus, in
Fig.~\ref{fig:RLF_1.4}, we also show the RLFs obtained by applying the observed
$L_{1.4~\rmn{GHz}}-L_{\rmn{X,bol}}$ and $L_{1.4~\rmn{GHz}}-Y_{\rmn{SZ}}$
relations (as in Fig.~\ref{fig:PLSZ}) to our $z = 0.2$ mock catalog, which
employs our phenomenological gas model for $L_{\rmn{X,bol}}$ and $Y_{\rmn{SZ},
  500}$ of each cluster, respectively. Note that this procedure is {\em only}
applied to halos defined as radio-loud clusters (which are by definition
accounted for in the $L_{1.4~\rmn{GHz}}$ scaling relations) and we assume a
fraction of 1, 0.25, 0.1, 0.05 and 0.01 of radio-loud clusters. As evident from
Fig.~\ref{fig:RLF_1.4}, this differs for our model scaling relations: there we
also define a fraction of radio-loud clusters, but the radio-quiet population
also contributes to the RLF with an increasing fraction at low
luminosities. This is exemplified in the top left plot of
Fig.~\ref{fig:RLF_1.4}, which shows the contribution of radio-quiet and loud
populations to the total RLF, assuming a fraction of 0.25 and 0.01 of radio-loud
objects.

In Appendix~\ref{app:D}, we make an attempt to construct an RLF from existing
X-ray flux-limited radio surveys. Of the few existing studies, we select the
cluster radio survey done with the National Radio Astronomy Observatory (NRAO)
Very Large Array (VLA) sky survey (NVSS) at $1.4$~GHz of
\cite{1999NewA....4..141G} and the survey with the Giant Metrewave Radio
Telescope (GMRT) at $610$~MHz by \cite{VenturiGMRT_1,VenturiGMRT_2}. For the
latter, we can also construct an RLF at 1.4~GHz using the corresponding RH
follow-up measurements. The fractions of radio-loud clusters are about 0.06,
0.18 and 0.24 for the NVSS 1.4~GHz, GMRT 610~MHz and GMRT 1.4~GHz samples,
respectively. As explained in Appendix~\ref{app:D}, we use the 1.4~GHz NVSS RLF
(with a median redshift of $z \approx 0.18$) as observational reference for our
comparisons. We conclude that the observational determinations of the RLF is not
very robust at this stage; the very different fractions of radio-loud clusters
found in the different studies is one indicator of this. Recently,
  \cite{2013arXiv1306.3102K} found only one additional radio mini-halo in an
  \emph{extended} GMRT survey. This does not significantly increase the
  statistics of RLF studies with respect to the sample of
  \cite{VenturiGMRT_1,VenturiGMRT_2}. We therefore decided to keep the 1.4~GHz
  NVSS RLF as our observational reference.
  
Generally, there is fair agreement between the NVSS RLF and both our modeled RLF
and the RLFs based on observational scaling relations, particularly for
radio-loud fractions between 10 per cent and 1 per cent. In particular, we
verified that the cumulative number of RHs above a certain flux limit of the
NVSS survey is well matched by the case of a radio-loud fraction of 10 per cent,
which will be used in the following section.  The RLF obtained from the
$L_{1.4~\rmn{GHz}}-L_{\rmn{X,bol}}$ relation differs from the NVSS RLF at high
luminosities, presumably caused by the large observed scatter.  On the other
side, the RLF obtained from the $L_{1.4~\rmn{GHz}}-Y_{\rmn{SZ}}$ relation
matches the NVSS result better. These results need to be consolidated by RLFs
corrected for flux-incompleteness and simulations of larger cosmological volumes
that are more complete at the high-mass end. Figure~\ref{fig:RLF_1.4}
demonstrates that it will be difficult to discriminate between different
scenarios at high radio luminosities (or equivalently masses). Indeed, in the
bottom right panel of Fig.~\ref{fig:RLF_1.4} we compare our RLF and the RLFs
based on observational scaling relations for a 10 per cent fraction of
radio-loud clusters to the NVSS RLF. This suggests that the low-luminosity
(low-mass) clusters will be the most useful in disentangling between different
models. This emphasizes the importance of conducting homogeneous, well
controlled surveys of RHs with the Jansky VLA, ASKAP \citep{2011PASA...28..215N}
and APERTIF \citep{2012JApA..tmp...34R} at 1.4 GHz and LOFAR at lower
frequencies. Since the latter has already started to take data, it is extremely
timely to present RLF predictions in this wavelength regime for our extended
hadronic model, which we will do next.

\subsection{Low-frequency predictions at 120~MHz}

\begin{figure*} 
\centering
\includegraphics[width=0.5\textwidth,height=0.305\textheight]{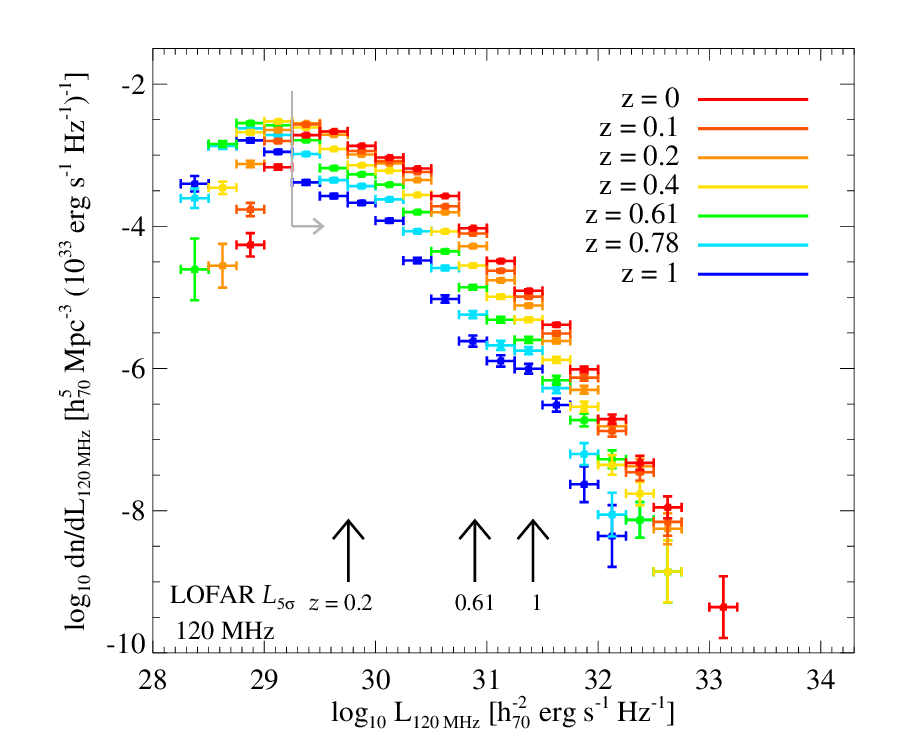}
\includegraphics[width=0.47\textwidth,height=0.3\textheight]{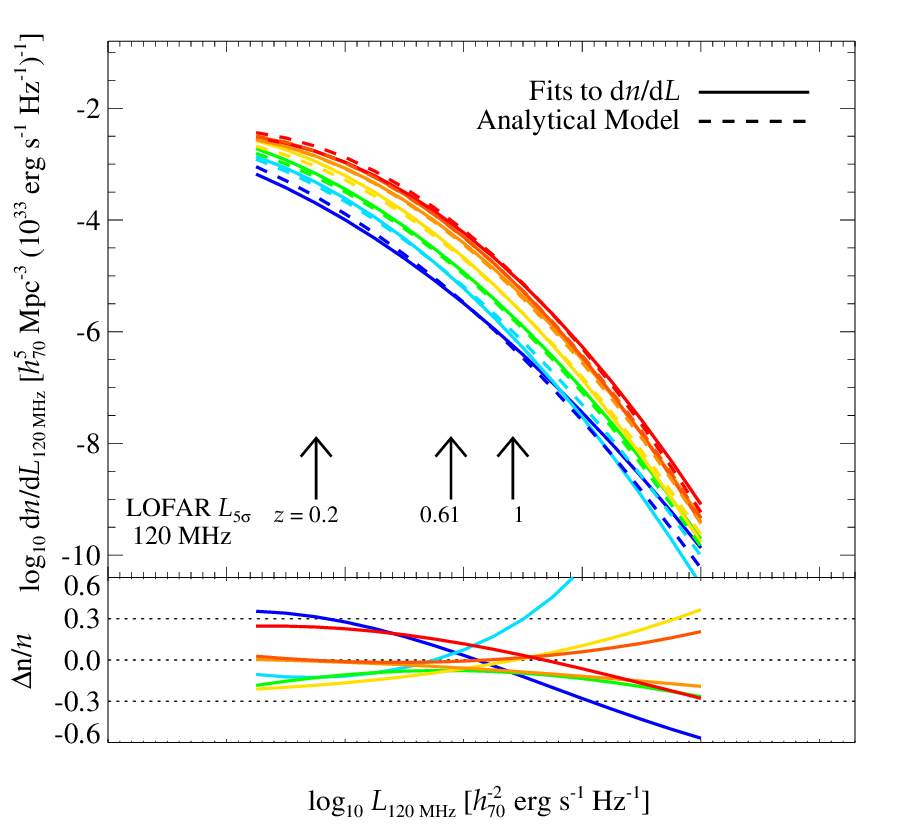}
\includegraphics[width=0.5\textwidth,height=0.305\textheight]{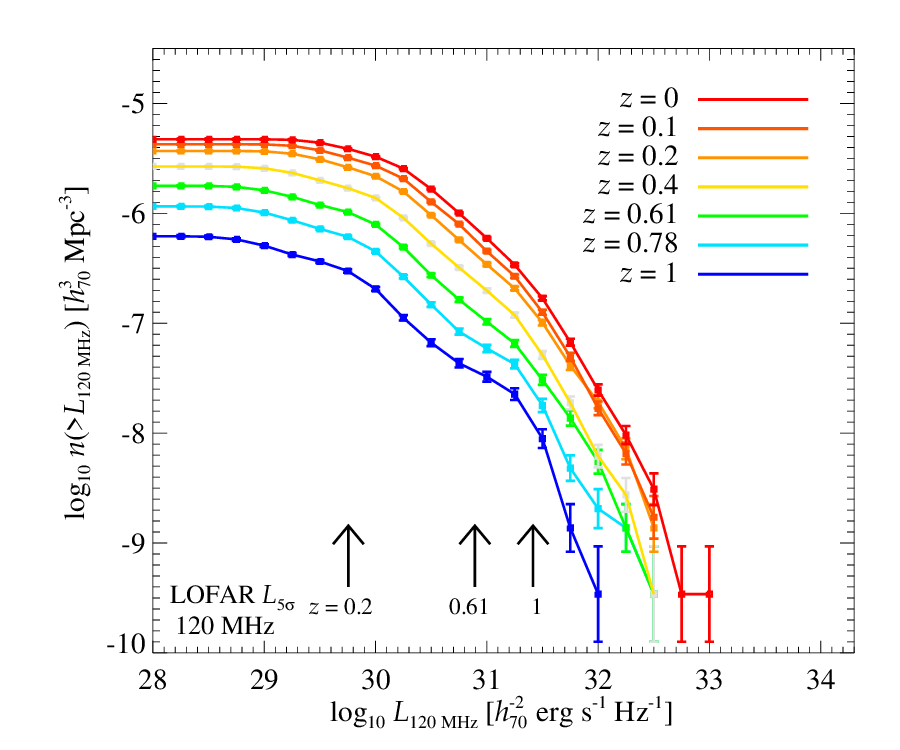}
\includegraphics[width=0.47\textwidth,height=0.3\textheight]{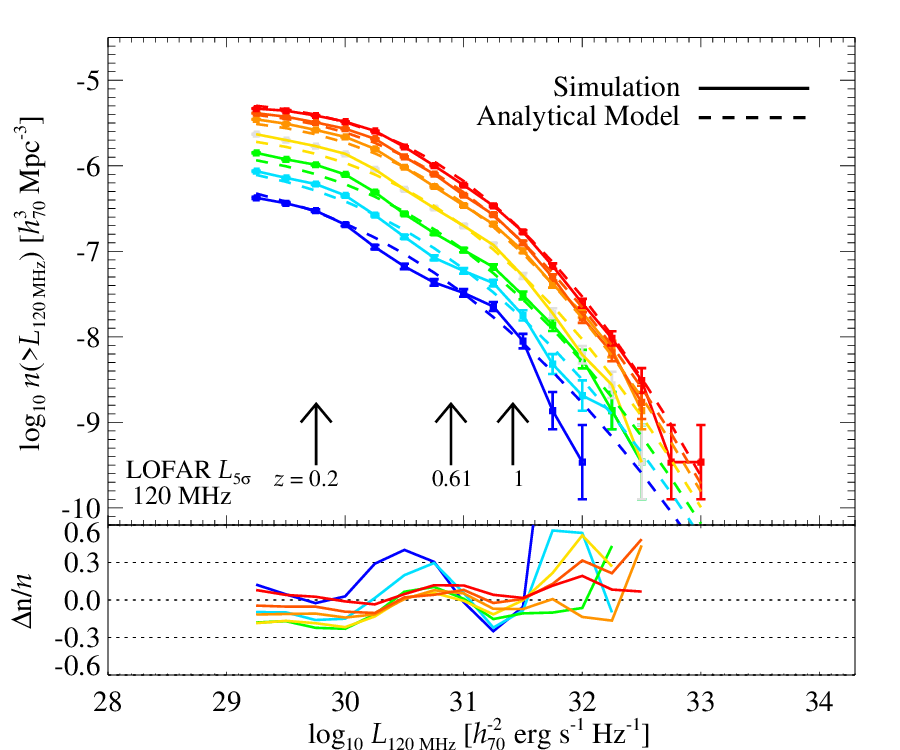}
\caption{RH luminosity function $n(L,z)$ at 120~MHz (top left panel) and
  cumulative number density of RHs $n(>L,z)$ (bottom left panel) at different
  redshifts $z$ (color coded) for the model realization described in
  Section~\ref{sec:4} with a 10 per cent fraction of radio-loud clusters. To
  obtain an analytical model for $n(L,z)$, we fit the logarithm of the RLF at
  each $z$ with a second-order polynomial and constrain the evolution of the
  three free parameters to follow a linear function in $1+z$ (see main text for
  details). The right panels show the comparison of the RLF fits (top) and the
  cumulative number density of the MultiDark samples (bottom) to the constrained
  analytical model.  The bottom panels on the right-hand side show the relative
  differences $\Delta n / n = (n_{\rmn{analytical}} -
  n_{\rmn{fit}})/n_{\rmn{fit}}$ (top) and $\Delta n / n = (n_{\rmn{analytical}}
  - n_{\rmn{simulation}})/n_{\rmn{simulation}}$ (bottom).  Additionally shown is
  the LOFAR Tier~1 \emph{point-source} flux limit of
  $F_{5\sigma}^{\rmn{PS}}=0.5$~mJy \citep{2012JApA..tmp...34R} converted to a
  luminosity limit at a given redshift. Horizontal error bars represent the mass
  bins while the vertical error bars are Poissonian uncertainties.  The light
  gray line marked by the arrow estimates the incompleteness limit owing to the
  adopted low-mass cut (in the construction of the mock cluster catalog, see
  Paper~I) and the scatter in the halo luminosities.}
\label{fig:RLF_120}
\end{figure*}

In Fig.~\ref{fig:RLF_120}, we show our model predictions at 120~MHz obtained
with the same representative realization of our model as in Section~\ref{sec:4},
with 10 per cent radio-loud clusters at all redshifts. We show both the
differential RLF (top left panel) and the cumulative number density (bottom left
panel) at different redshifts (corresponding to different the MultiDark
snapshots in Table~1 of Paper~I).  We note that the redshift evolution is almost
entirely due to the $B^2/(B^2+B_{\rmn{CMB}}^2)$ factor of
equation~(\ref{eq:jnu}) since $B_\rmn{CMB}\propto (1+z)^2$.  Our imposed mass
cutoff of $M_{200}=10^{14}\,h^{-1}\,M_\odot$, which has been adopted to reliably
model the cluster gas distribution (Paper~I), translates into a luminosity
cutoff. This causes the differential luminosity function to turn over at the
low-luminosity end and artificially flattens the slope already at sightly higher
luminosities than the luminosity maximum that indicates our formal
incompleteness limit.  Note that calibrating our model to 1.4~GHz
  observations may cause an over- or underestimate of the number of
  low-frequency halos as a result of intrinsic spectral flattening towards low
  frequencies (e.g., due to CR transport) or steeper leptonic spectra in
  comparison to the hadronic component in the high-mass radio-loud NCCC
  population, respectively.

Additionally shown in Fig.~\ref{fig:RLF_120} is the expected LOFAR Tier~1
\emph{point-source} flux limit of $F_{5\sigma}^{\rmn{PS}}=0.5$~mJy
\citep{2012JApA..tmp...34R} converted to a luminosity limit at a few
representative redshifts. This flux limit is clearly an underestimate for nearby
RHs, which extend over angular scales $\sim1$~deg, as e.g., in the case of the
Coma radio halo. In order to make more reliable predictions, we will
  calculate the RH flux limit with equation~(10) of \cite{2010A&A...509A..68C}
  that is based on the assumption that RHs emit about half of their total radio
  flux within their half radius. In our sample, the typical radius within which
  half of the radio flux is emitted is $R_{500}/4$, and we require that the flux
  within this radius is higher than $F_{5\sigma}^{\rmn{PS}}$. The median
  $R_{500}$ of our sample is about $0.8$~$h_{70}^{-1}$~Mpc at all
  redshifts. This translates to a flux limit of about $48, 14, 3.5$ and
  $2.5$~mJy at $z = 0.1, 0.2, 0.6$ and $1$, respectively.

\begin{figure*} 
\centering
\includegraphics[width=0.48\textwidth]{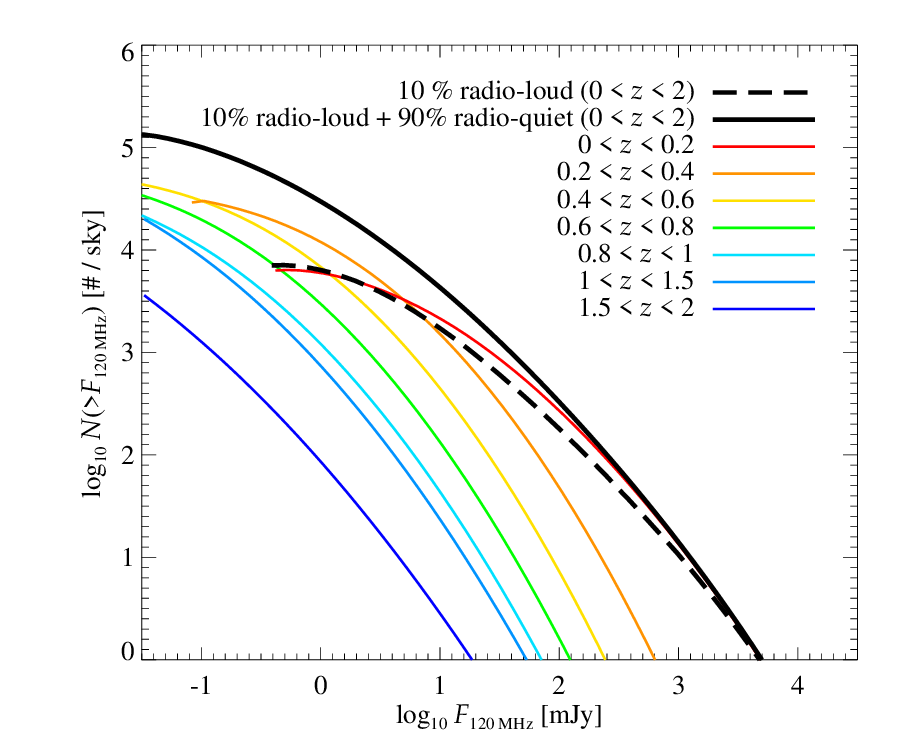}
\includegraphics[width=0.48\textwidth]{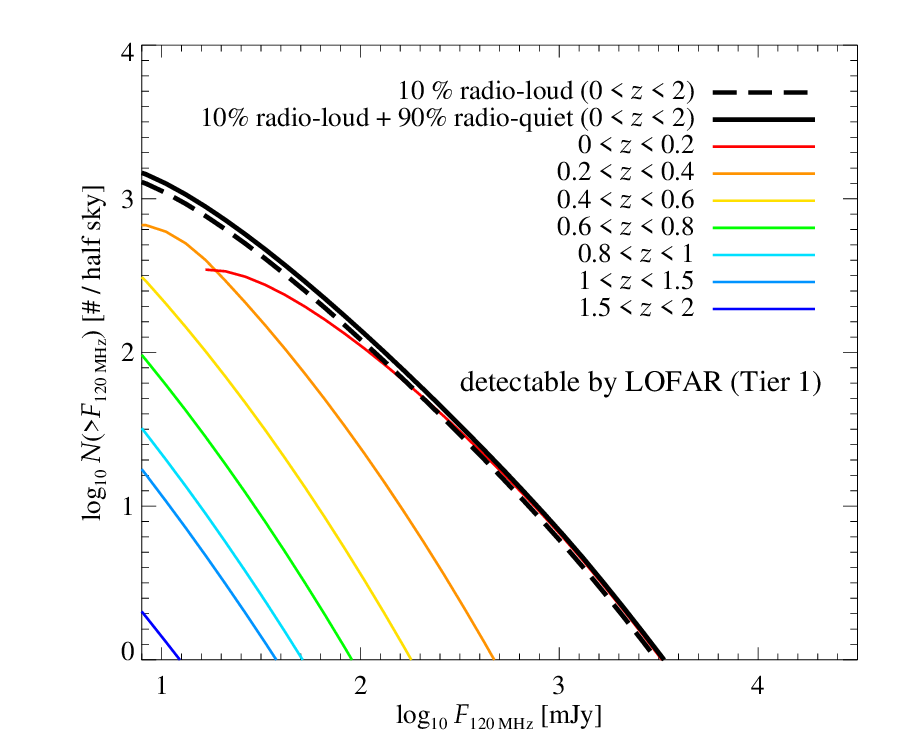}
\caption{Cumulative number of RHs above a certain flux limit in an all-sky
  survey at 120~MHz. We show the result of the model realization described in
  Section~\ref{sec:4} using all clusters and adopting a fraction of 10 per cent of
  radio-loud clusters (black solid line). Additionally, we show the result
  obtained by using the 10 per cent radio-loud clusters \emph{only} (black dashed
  line). We also show the differential contribution to the RLF in redshift
  slices. Note that the number of (detectable) RHs would be dramatically reduced
  by the presence of a break in the model at some low luminosity-scale, or a
  mass-dependence of the model parameters causing the RH luminosities to
  decrease at low masses.  \emph{Left.} Total number of RHs in the sky.
  \emph{Right.} Number of detectable RHs by the LOFAR Tier~1 survey considering
  its sky coverage (half sky) and adopting a realistic flux limit corresponding
  to different angular source extensions at different redshifts (see main text for details).
}
\label{fig:RLF_120_flux}
\end{figure*}

To derive low-frequency flux functions, we construct an analytical model for the
evolving RLF. We fit the 120~MHz RLF at different redshifts with a second-order
polynomial of the form $\log_{10} \rmn{d}n/\rmn{d}L_{120~\rmn{MHz}} = A_{0} +
A_{1}~\log_{10} L_{120~\rmn{MHz}} + A_{2}~(\log_{10} L_{120~\rmn{MHz}})^{2}$.
All luminosities are measured in units of $h_{70}^{-2}
\rmn{erg~s}^{-1}\,\rmn{Hz}^{-1}$ and (comoving) number densities in units of
$h_{70}^{3}\,\rmn{Mpc}^{-3}$. We consider only luminosities with $\log_{10}
L_{120~\rmn{MHz}} \geq 29.25$ to exclude the turn-over at low luminosities
caused by incompleteness. To obtain an analytical model for $n(L,z)$, we
constrain the evolution of the three free parameters $A_i$ to follow a linear
function in $1+z$, i.e., $A_{i} = A_{i,0} + A_{i,1}~(1+z)$.\footnote{The values
  of these parameters are $A_{0,0} = -484.74$, $A_{0,1} = 141.50$, $A_{1,0} =
  32.66$, $A_{1,1} = -9.07$, $A_{2,0} = -0.55$ and $A_{2,1}$ = 0.14.} In the
right panels of Fig.~\ref{fig:RLF_120}, we compare the RLF fits (top) and the
cumulative number density in the simulation (bottom) to the analytical model. In
particular the analytics matches the simulation well except for high
luminosities and high redshifts where small number statistics explains the
deviations.

This analytical model describes the number of RHs expected in our model per unit
luminosity and per unit comoving volume $V_{\rmn{c}}$, i.e., $\rmn{d}^2N(L,z)
/\rmn{d}V_{\rmn{c}}\rmn{d}L$. Hence the cumulative number of RHs above a given flux
limit $F$ is given by the integral
\begin{equation}
N(>F)  =  \int_{z_1}^{z_2} \int_{L(F)}^{\infty} 
\frac{\rmn{d}^2N(L,z)}{\rmn{d}V_{\rmn{c}}\rmn{d}L}\,
\frac{\rmn{d}V_{\rmn{c}}}{\rmn{d}z}\, \rmn{d}z\, \rmn{d}L ,
\label{eq:NtotRH}
\end{equation}
where $L(F) = 4 \pi D(z)^2 F$ and $D(z)$ is the luminosity distance to an RH at
redshift $z$.  The result is shown in the left panel of
Fig.~\ref{fig:RLF_120_flux} for the model realization described in
Section~\ref{sec:4} with a 10 per cent fraction of radio-loud clusters (black
solid line). We limit the integral to luminosities $\log_{10} L_{120~\rmn{MHz}}
\geq 29.25$. Our redshift integration extends from $z_{1} = 0.018$, the redshift
of the closest known RH in Perseus, to $z_{2} = 2$. As shown in
Fig.~\ref{fig:RLF_120_flux}, already redshifts $z\gtrsim1$ do not significantly
contribute to the flux function. Additionally, there are large theoretical
uncertainties since our gas model is not calibrated for these redshifts (and
implied low-mass range) and very little is observationally known about diffuse
radio emission on group scales in particularly at these redshifts, which
motivates our upper redshift limit. Note also that at these high redshifts,
  the analytical fit to the evolving RLF slightly overproduces the simulation
  number counts.

Figure~\ref{fig:RLF_120_flux} shows the contribution of different redshift
slices to the total flux function. We contrast this to the flux function using
\emph{only} the subsample of 10 per cent radio-loud clusters (black dashed line). This
was obtained by constructing the corresponding RLF and repeating the steps above
in building an analytical model, however, discarding luminosities $\log_{10}
L_{120~\rmn{MHz}} \leq 30.75$ in the integration.

In the right panel of Fig.~\ref{fig:RLF_120_flux}, we show the total number of
RHs that would be \emph{detectable} by the LOFAR Tier~1 survey, where its sky
coverage (of about half the entire sky) and the signal degradation due to source
extensions of close-by RHs is taken into account. The latter is calculated with
equation~(\ref{eq:NtotRH}) and adopting $F = F_{\rmn{min}}$ where
$F_{\rmn{min}}$ is given by equation~(10) of \cite{2010A&A...509A..68C} as
explained above. Comparing the left and right panel of Fig.~\ref{fig:RLF_120_flux}
elucidates the critical impact of the detection threshold on the number observable RHs.
A detailed characterization of the instrumental response is needed in order
to obtain more precise estimations.

The LOFAR Tier~1 survey at 120~MHz should be able to detect about $1400$
  clusters hosting radio giant and mini halos, considering the hadronic
  component only. We refer the reader to \cite{2010A&A...509A..68C} for
  predictions for giant halos in the turbulent re-acceleration scenario. We note
  that those are complementary, because they address a larger mass scale with
  respect to ours, which are limited by the volume of the MultiDark simulation
  (see Paper~I).

The precise number of detections depends strongly on the underlying
assumptions. There are two main uncertainties in our model: the fraction of
radio-loud to radio-quiet clusters and the corresponding luminosities as well as
our assumed RH modeling in low-mass clusters (which are not yet known to host
RHs). The fraction of radio-loud to radio-quiet clusters is determined from a
given (degenerate) set of model parameters that include $\gamma_{\rmn{th}}$,
$B_{0}$, $\alpha_{B}$, $g_{\rmn{CR}}$, or equivalently $X_{CR}$. The fraction of
radio-loud clusters mostly affects the number of medium-to-high luminosity RHs
(as can be seen from the 1.4~GHz RLFs in the top left panel of
Fig.~\ref{fig:RLF_1.4}). The total number of RHs is dominated by low-luminosity
RHs. While this may suggest that the radio-loud fraction is of minor importance
for the number of detectable RHs, the opposite is the case. Because of the flux
limit, only the most luminous clusters at each redshift are observable so that
the total number of detectable RHs scales almost linearly with the radio-loud
fraction.

We caution that our predicted total number of (detectable) RHs depends on the
ability to extrapolate the observed and modeled scalings down to our adopted
mass limit of $M_{200}\approx1.4\times10^{14}$~$h_{70}^{-1}$~M$_{\odot}$. If the
underlying physics imprinted a characteristic scale into the CR transport or
magnetic field distribution, this would manifest itself as a break in the radio
luminosity scaling relations and dramatically reduce (or even increase) the
number of expected RHs.  While this does not interfere with previous
(high-frequency) measurements at high luminosities, this may of critical
importance for future, more sensitive (low-frequency) measurements of
low-redshift clusters that probe the uncertain regime of diffuse radio emission
in low-mass clusters.

The reason for this lies in the steep halo mass function which ensures that that
clusters above a given cutoff (which is either physically motivated or
observationally realized through a survey flux limit) dominate the total number
of (detectable) RHs. Only if the radio luminosity scaling remains unaltered
below the survey flux limit, the steepness of the mass function ensures that
there will be more RHs scattered above the flux limit then below it. This is the
so-called Eddington bias that causes the inferred luminosities based on only the
detected sources to end up as an overestimate. Hence the presence of any
hypothetical break in the radio luminosity scaling, which is unconstrained by
current data, explains the largely varying model predictions in the recent
literature, which vary from a few to hundreds observable RHs
\citep{2010A&A...509A..68C,2012arXiv1210.1020C,2011arXiv1110.2786S} to thousands 
of detectable RHs in future surveys \citep{2002A&A...396...83E}. 

Another relevant issue in such surveys is the identification of RHs and their
hosting clusters (see also \citealp{2010A&A...509A..68C}). RHs constitute a
small part of the entire (diffuse as well as apparent point-like) radio source
population and therefore need to be distinguished from the emission produced by
other sources. A good approach will be to cross-correlate the radio maps with
high-sensitivity X-ray surveys such as that by the future \emph{e}ROSITA mission, which
is expected to detect around $10^{5}$ clusters up to redshift $z \approx 1.3$
(e.g., \citealp{2011MSAIS..17..159C}).

Our results show the prospects of the LOFAR survey and other future radio
instruments in determining the RH properties over a broad range of luminosities.
This should permit a robust determination of the number of clusters hosting RHs 
at a given luminosity (mass) and to carefully asses completeness issues. This will 
be extremely helpful in elucidating the RH generation mechanism, in determining the 
viable parameter space for the hadronic model and in establishing the precise role of 
the hadronic contribution to the total radio emission of merging clusters. 
In particular, the comparison with the predictions of the turbulent re-acceleration 
scenario by \cite{2010A&A...509A..68C,2012arXiv1210.1020C}, where a significantly
smaller number of objects is expected to be detected, will eventually help in disentangling 
different models and in identifying the importance of the hadronic component.

\section{Conclusions}
\label{sec:6}

This paper aims at scrutinizing the hadronic model for giant and mini radio
halos. Our phenomenological modeling of the CR distribution is guided by
cosmological cluster simulations and theoretical considerations of microscopic
CR transport. It is designed to efficiently and simultaneously model the RH
surface brightness emission, RH scaling relations and luminosity
functions. Within the hadronic scenario, the interplay of CR advection and
streaming appears to be crucial to match observed RH distributions as a function
of SZ flux as well as to explain the bimodality of radio-loud and radio-quiet
clusters at a fixed X-ray luminosity. However, the spatial extend of giant radio
halos is difficult to accomplish within the hadronic framework, especially for
the Coma halo at low frequencies \citep{2012arXiv1207.3025B}. This calls for a
revision of purely hadronic models for giant radio halos.

To construct an \emph{extended} model for the CR distribution in clusters, we
adopt the universal spatial and spectral CR distribution found in hydrodynamic
cosmological simulations of cluster formation \citep{2010MNRAS.409..449P}. Since
these simulations only follow the macroscopic, advective CR transport, we
additionally account for microscopic CR transport processes
\citep{2011A&A...527A..99E,2013arXiv1303.4746W}. While turbulently-driven CR 
advection can lead to centrally enhanced CR profiles, CR propagation in the form 
of CR streaming and diffusion produces flatter CR profiles, which should be realized 
for decaying cluster turbulence. In our model, we introduce a CR propagation parameter
$\gamma_{\rmn{tu}}$ that is the ratio of the CR streaming-to-advection time
scale. This parameter allows us to effectively switch the regimes where either
process dominates the CR transport and to explore different turbulent states of
clusters.

This enables us to model the radio surface brightness profiles of giant radio
halos (as exemplified in Coma and Abell~2163) as well as of radio mini-halos (in
Perseus and Ophiuchus) at 1.4~GHz. We find an excellent match to mini halos over
a wide range of parameter choices, rendering the hadronic model as an attractive
explanation for mini halos. However, in order to match the extended surface
brightness profiles of giant halos at {\em high} frequencies (1.4 GHz), the
hadronic model would require flat CR profiles for magnetic field configurations
favored by Faraday rotation measurements. These flat CR profiles should only be
realized through CR streaming transport in {\em relaxed} clusters, which appears
to be in conflict with the observation that giant radio halos are hosted by
merging {\em turbulent} clusters.\footnote{Note, however, that
    \cite{2013arXiv1303.4746W} arrive at a different conclusion finding that the
    increase of turbulence promotes outward streaming more than inward
    advection, therefore allowing flat CR distributions in turbulent clusters.}
  Moreover, the hadronic model fails to explain the emission in the outer
  parts of the Coma halo at 352~MHz where not even the extreme case of a flat CR
  profile is sufficient to explain the emission as of purely hadronic origin.
This motivates us to suggest the following new \emph{hybrid} hadronic-leptonic
scenario.
\begin{enumerate}
\item Radio mini halos are primarily of hadronic origin.  
\item Giant radio halos experience a transition from the central hadronic
  emission component to a dominantly leptonic emission component in the outer
  halo that is due to Fermi I or II re-acceleration of fossil or hadronically
  produced electrons.
\item Steep spectrum radio sources are mainly of leptonic origin.
\end{enumerate}
This scenario would imply an increased spectral and morphological variability in
leptonically dominated emission regions because of the intermittency and
relative inefficiency of the corresponding re-acceleration processes (Fermi I
acceleration at weak intra-cluster shocks or Fermi II acceleration at plasma
waves). In particular, it implies a spectral steepening from the hadronic to the
leptonic component, since the long-lived CR protons are dominantly accelerated
by stronger formation shocks during the gas assembling history onto a cluster,
which causes a harder spectrum in comparison to the softer leptonic component.
We checked that for parameter ranges that provide acceptable matches to the
radio profiles, the resulting gamma-ray emission from the decay of neutral
pions---an inevitable by-product in hadronic CR interactions---is below
observational gamma-ray upper limits provided by {\em Fermi} and imaging
atmospheric Cherenkov telescopes.

To address the hadronic RH statistics, we apply our extended CR model 
to a cosmologically complete mock cluster catalog built from the MultiDark
$N$-body simulation in a companion paper (Paper~I). We select a representative 
realization of our extended CR model and compare it with existing radio scaling relations. 
Because of CR transport and the different gas-density scalings of the X-ray luminosity,
$L_{\rmn{X}}\propto\int \rho_\rmn{gas}^2 \dd V$, and the Sunyaev-Zel'dovich
flux, $Y\propto\int \rho_{\rmn{gas}} k_{\rmn{B}} T \dd V$, our model is able to
simultaneously reproduce the observed bimodality of radio-loud and radio-quiet
clusters at the same $L_{\rmn{X}}$ as well as the unimodal distribution of
radio-halo luminosity versus $Y$; thereby suggesting a physical solution to this
apparent contradiction. We caution however, that some parameters in our model
are degenerate with respect to the resulting radio luminosity and radio emission
profiles, in particular $\gamma_{\rmn{tu}}$ and $ \alpha_{\rmn{B}}$ (our rate of decline
of the magnetic field toward the cluster outskirts). Multi-frequency data will
be needed to better constrain these parameters and to break these degeneracies.

Assuming a redshift-independent fraction of 10 per cent radio-loud
clusters, we demonstrate that our model matches the NVSS RH luminosity function
(RLF). However, the high-luminosity tail of our model RLF is subject to cosmic
variance because of the comparably small simulation volume of $1~h^{-3}
\,\rmn{Gpc}^3$. Interestingly, the RLF derived from the
$L_{1.4~\rmn{GHz}}-L_{\rmn{X,bol}}$ relation differs at high radio luminosities
from the NVSS RLF; possibly because of selection and incompleteness effects that
are not fully taken into account. The comparison between different RLFs suggests
that the low-luminosity (low-mass) regime is the most promising place to
differentiate between various models.

It is expected that the next-generation of low-frequency radio surveys will
probe this regime. Hence, we we make prediction for the LOFAR cluster survey, in
particular, we compute the 120~MHz RLF and the cumulative RH number
density. Given our assumptions, we would expect the LOFAR Tier~1 survey at
120~MHz to detect about $1400\times (\rmn{\emph{radio-loud~fraction}/0.1})$ 
hadronically-generated radio giant and mini halos. 
We caution that the precise number depends strongly on the underling assumptions. 
In particular, we assume that the model parameters can be extrapolated down to cluster 
masses of about $M_{200}\approx1.4\times10^{14}$~$h_{70}^{-1}$~M$_{\odot}$ without 
any break in the radio scaling relation that would indicate additional scales in the physics.
Most of the RHs in our sample lie at low masses and thus at low luminosities
that are unconstrained by current observations. If, e.g., the magnetic field
and/or the CR distribution in clusters are not statistically self-similar and
would exhibit much reduced strength/number density on group scales, our
predictions for the detectable number of RHs would be dramatically reduced.

This demonstrates the potential of LOFAR, and other next-generation
high-sensitivity radio instruments such as APERTIF, ASKAP, EVLA and SKA, in
determining the RLF properties. In combination with future X-ray missions like
\emph{e}ROSITA, this should yield a robust determination of the number of
clusters hosting RHs at a given luminosity (mass) and thus elucidate the
relation of the radio emission with the dynamical state of a cluster.
This is crucial in order to understand the RH generation mechanism and to
establish the precise role of the hadronic contribution in clusters.

We have constructed a model for the ICM (in Paper~I) and CR distributions 
in galaxy clusters that enables us to provide a cosmologically complete multi-frequency mock 
catalog for the \mbox{(non-)thermal} cluster emission at different redshifts. We make these 
catalogs publicly and freely available on-line through the MultiDark database (\emph{www.multidark.org}, \citealp{2013AN....334..691R}). 
We hope that the community can make valuable use of these catalogues in
synergy with the future radio, X-ray and gamma-ray data.

\section*{Acknowledgments}
We thank the anonymous referee for the useful comments.
We thank Anders Pinzke for many useful discussions and Lawrence Rudnick and Shea
Brown for a reanalysis of their 352~MHz radio map of Coma, and for useful
discussions. We thank Matteo Murgia for kindly providing the radio surface
brightness profiles of Ophiuchucs and Abell~2163 that have been recomputed with
respect to the Reiprich \& B\"{o}ringher (2002) cluster positions, and Wolfgang
Reich for providing the 1.4~GHz radio map of Coma.  We also thank Miguel-Angel
Perez Torres and Huub R{\"o}ttgering for the useful discussions.  Finally, we
thank the MultiDark database people, in particular Adrian Partl and Kristin
Riebe.  F.Z.{\ }acknowledges the CSIC financial support as a JAE-Predoc grant of
the program ``Junta para la Ampliaci\'on de Estudios'' co-financed by the FSE.
F.Z.{\ }acknowledges the hospitality of the Leiden Observatory during his stay.
F.Z.{\ }and F.P.{\ } thank the support of the Spanish MICINN's
Consolider-Ingenio 2010 Programme under grant MultiDark CSD2009-00064, AYA10-21231. 
C.P.{\ }gratefully acknowledges financial support of the Klaus Tschira Foundation. The
MultiDark Database used in this paper and the web application providing online
access to it were constructed as part of the activities of the German
Astrophysical Virtual Observatory as result of a collaboration between the
Leibniz-Institute for Astrophysics Potsdam (AIP) and the Spanish MultiDark
Consolider Project CSD2009-00064, AYA10-21231. The Bolshoi and MultiDark simulations 
were run on the NASA's Pleiades supercomputer at the NASA Ames Research Center.

\bibliographystyle{mnras}
\bibliography{bib_file}

\begin{appendix}

\section{Cosmic Ray Modeling}
\label{app:B}

Here, we describe in detail how our \emph{extended} model for the CR distribution
in galaxy clusters of Section~\ref{sec:2.3} is constructed by generalizing the
analytical results by \cite{2011A&A...527A..99E}.

As anticipated in Section~\ref{sec:2.3}, when advection dominates the CR
transport, the CR normalization can be expressed as in
equation~(\ref{eq:Csimple_1}). However, when CR streaming and diffusion
dominates, the CR distribution is modified and flattens considerably. This can
be shown analytically by solving the continuity equation for CRs and obtaining
the CR density profile, $\rho_{\rmn{CR}}$, of
equation~(\ref{eg:rhoCR_1}). Assuming $P_{\rmn{th}}(R)/P_{\rmn{th},0}=n_{\rmn{e}}(R)/n_{0}$, i.e.,
neglecting the temperature dependence, and adopting a standard $\beta$-profile
for the electron density,
\begin{equation}
n_{\rmn{e}} = n_{0} \left( 1+\frac{R^{2}}{R_{\rmn{c}}^{2}} \right)^{-\frac{3\beta_{\rmn{cl}}}{2}} \, ,
\label{eq:beta_profile}
\end{equation}
\cite{2011A&A...527A..99E} find that the solution of equation~(\ref{eg:rhoCR_1}) is physical only 
for $\rho_{\rmn{CR}}$ within the radial range $R_{-} < R < R_{+}$ with
\begin{equation}
R_{\pm} = \frac{3\beta_{\rmn{cl}}}{2\gamma}R_{*}\left(1\pm\sqrt{1-\left(\frac{2R_{\rmn{c}}\gamma}{3\beta_{\rmn{cl}}R_{*}}\right)^{2}}\right) \, ;
\label{eq:Rpm}
\end{equation} 
while it is non-stationary outside these radii. In these regions, the authors
suggest to set $\rho_{\rmn{CR}}(R) = \rho_{\rm{CR}}(R_{\pm})$ for $R > R_{+}$
and $R < R_{-}$, respectively. \cite{2011A&A...527A..99E} obtain the profile for
the CR normalization,
$C(R)=C_{0}(\rho_{\rmn{CR}}(R)/\rho_{\rmn{CR},0})^{\beta_{\rmn{CR}}}$, as
\begin{equation}
C(R) = C_{0}\left( 1+ \frac{R^{2}}{R_{C}^{2}} \right)^{-\beta_{\rmn{c}}} \rmn{exp}\left( {\frac{R}{R_{*}}\beta_{\rmn{CR}}} \right)
\label{eq:Ctransport}
\end{equation} 
for $R_{-}<R<R_{+}$, where
$\beta_{\rmn{c}}=3\beta_{\rmn{cl}}~\beta_{\rmn{CR}}/2\gamma$, and $C(R) =
C(R_{\pm})$ for $R<R_{-}$ and $R>R_{+}$, respectively. In this way, different CR
transport cases are parametrized through $\gamma_{\rmn{tu}}$.  A high value of
$\gamma_{\rmn{tu}}$ characterizes the advection-dominated case, while the CR
profile is flat for $\gamma_{\rmn{tu}} \sim1$. We refer the reader to
\cite{2011A&A...527A..99E} for an extensive discussion.

\begin{figure}
\centering
\includegraphics[width=0.5\textwidth]{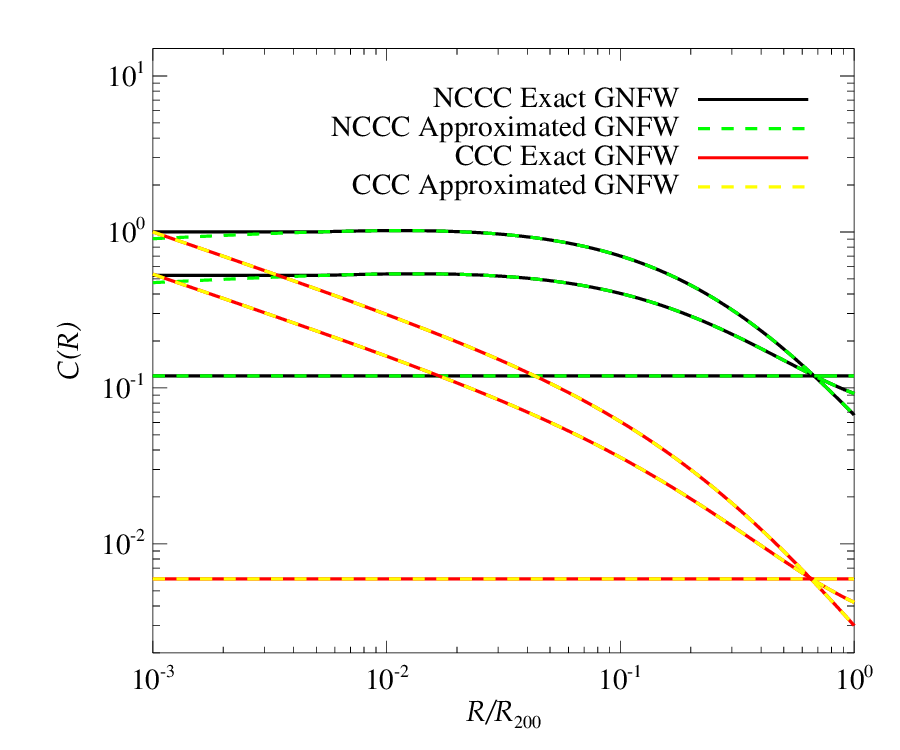}
\caption{Comparison of the exact and approximate solutions for $C(R)$ in the
  case of our GNFW profile. In the latter case, we use the formulae by
  \protect\cite{2011A&A...527A..99E} but adopt $P(R)/P_{0}=n_{\rmn{e,GNFW}}(R)/n_{0}$
  for this comparison only, i.e., we assume here an isothermal ICM. We show the GNFW
  profiles for NCCCs and CCCs, as derived in Paper~I. For each of
  these profile classes, we show three different cases of
  $\gamma_{\rmn{tu}}=100$, 10, and 1 (top to bottom). We normalize $C(R)$ for
  the advection dominated case ($\gamma_{\rmn{tu}}=100$) at $R=10^{-3}R_{200}$
  and and require CR number conservation during CR streaming. Note that the
  value for $C(10^{-3}R_{200},\gamma_{\rmn{tu}}=100)$ in the CCC case is
  identical for the exact solution and its approximation, while there is a small
  difference of about 9 per cent in the NCCC case. We adopt $\alpha=2.3$.}
\label{fig:REXexactVSfake}
\end{figure}

We want to extend this result to account for (i) our GNFW gas profiles of
Paper~I, (ii) the universal temperature drop in the cluster
outskirts, and (iii) merge it with the universal cluster mass-scaling of the CR
normalization, $\tilde{C}$, obtained from hydrodynamical simulations
\citep{2010MNRAS.409..449P}.  Therefore, we adopt the extended profile,
$C_{\rm{extended}} = \tilde{C}(R) (\rho_{\rmn{gas}}(R)/m_\rmn{p}) (T(R)/T_0)$, of
equation~(\ref{eq:Cf}). We note that for such a choice, there does not any more
exist an analytical solution to equation~(\ref{eg:rhoCR_1}) as in
\cite{2011A&A...527A..99E}. This results in a 5-order equation and a numerical
solution would not be of practical use for every cluster in our large MultiDark
sample. For simplicity, we adopt the formulae above also for our extended
model. I.e., we adopt
$C(R)=C_{0}(\rho_{\rmn{CR}}(R)/\rho_{\rmn{CR},0})^{\beta_{\rmn{CR}}}$ within
$R_{\pm}$ of equation~(\ref{eq:Rpm}), with $\rho_{\rmn{CR}}$ defined by
equation~(\ref{eg:rhoCR_1}) where $C_{\rmn{extended}}$ enters through the
advective CR profile, $\eta(R)$, of equation~(\ref{eq:eta}), and $C(R) =
C(R_{\pm})$ for $R > R_{+}$ and $R < R_{-}$, respectively.  Note that, in our
formalism, $R_{\rmn{c}}$ of equation~(\ref{eq:Rpm}) becomes the characteristic
radius of our GNFW gas profile (see Paper~I), i.e., $R_{\rmn{c}} =
0.2 R_{500}$, and $\beta_{\rmn{cl}}=0.8$ (we checked that varying the value of
$\beta_{\rmn{cl}}$ between 0.4 and 1.2 has no impact).  The relevant factors are
$\gamma_{\rmn{tu}}$ and the exponential factor of
equation~(\ref{eg:rhoCR_1}). As we will see in the following, the two radii
$R_{\pm}$ are not critically affected by the form of $\eta(R)$. Therefore,
despite the approximation, this approach captures the main CR transport effects.

Assuming our GNFW gas profile instead of a standard $\beta$-profile for the
electron density, so adopting $P(R)/P_{0}=n_{\rmn{e,GNFW}}(R)/n_{0}$, there
exist an exact analytical solution to equation~(\ref{eg:rhoCR_1}) as in
\cite{2011A&A...527A..99E}. In order to evaluate the systematic error that we
are introducing with the approach described above, in
Fig.~\ref{fig:REXexactVSfake}, we compare the \emph{exact} solution for $C(R)$
in the case of our GNFW profile with the \emph{approximate} solution where we
use the formulae by \cite{2011A&A...527A..99E}, and only substitute the electron
$\beta$-profile for our GNFW profiles (with $R_{\rmn{c}} = 0.2 R_{500}$,
$\beta_{\rmn{cl}}=0.8$). In this last case, we fix $R_{-}=10^{-3}R/R_{200}$ to
mimic the typical $R_{-}$ value of the exact solution, otherwise an unphysical
step feature would appear at $\leq10^{-2}R/R_{200}$. This latter approximation
is kept in our extended model, which however has no impact on the model surface
brightness and total luminosity. There is almost no difference between the two
cases, as clear from Fig.~\ref{fig:REXexactVSfake}. This shows that the approach
presented here to construct our extended model can be safely followed in order to
derive a fully working parametrization which captures the main CR transport
effects.


\section{Radio Emission}
\label{app:A}

The synchrotron emissivity $j_{\nu}$, at frequency $\nu$ and per
steradian, of a steady-state electron population where radiative cooling
balances injection from hadronic interactions (adapted from
\citealp{2008MNRAS.385.1211P} and \citealp{2011A&A...527A..99E})
is given by
\begin{equation}
j_{\nu}  =  A_\nu C(R) \rho_{\rmn{gas}}(R) 
\frac{\epsilon_{\rmn{B}}(R)}{\epsilon_{\rmn{B}}(R)+\epsilon_{\rmn{CMB}}} 
\left( \frac{\epsilon_{\rmn{B}}(R)}{\epsilon_{B_{\rmn{c}}}} \right)^{(\alpha-2)/4} \, ,
\label{eq:jnu}
\end{equation}
where $\epsilon_{\rmn{CMB}}$ is the energy density of the cosmic microwave 
background (CMB), $\epsilon_B=B^{2}/(8\pi)$ denotes the magnetic
energy density, and $B_{\rmn{c}} = \sqrt{ 8 \pi
  \epsilon_{B_{\rmn{c}}}} \simeq 31 \left( \nu/\rmn{GHz} \right)~\umu$G. 
The factor $A_{\nu}$ is given by: 
\begin{equation}
A_{\nu} = A_{\rmn{E_{synch}}} \frac{16^{2-\alpha_{\rmn{e}}}\sigma_{\rmn{pp}}m_{\rmn{e}}c^{2}}{(\alpha_{\rmn{e}}-2)\sigma_{\rmn{T}}\epsilon_{B_{\rmn{C}}}m_{\rmn{p}}}\left(\frac{m_{\rmn{p}}
}{m_{\rmn{e}}}\right)^{\alpha_{\rmn{e}}-2} \left(\frac{m_{\rmn{e}}c^{2}}{\rmn{GeV}}\right)^{\alpha_{\rmn{e}}-1} \, ,
\end{equation}
with
\begin{equation}
A_{\rmn{E_{synch}}} = \frac{\sqrt{3\pi}}{32\pi}\frac{B_{\rmn{c}}e^{3}}{m_{\rmn{e}}c^{2}}\frac{\alpha_{\rmn{e}}+\frac{7}{3}}{\alpha_{\rmn{e}}+1}\frac{\Gamma\left(\frac{3\alpha_{\rmn{e}}-1}{12}\right)\Gamma\left(\frac{3\alpha_{\rmn{e}}+7}{12}\right)\Gamma\left(\frac{\alpha_{\rmn{e}}+5}{4}\right)}{\Gamma\left(\frac{\alpha_{\rmn{e}}+7}{4}\right)} \, ,
\end{equation}
where $\alpha_{\rmn{e}}=\alpha+1$, the effective inelastic cross-section for
proton-proton interactions is $\sigma_{\rmn{pp}} =
32~(0.96+\rmn{e}^{4.42-2.4\alpha})$, and
$\Gamma$ is the Gamma-function
\citep{1965hmfw.book.....A}. $A_{\rmn{E_{synch}}}$ is given in units of erg, and
$A_{\nu}$ is given in units of erg~cm$^{3}$~g$^{-1}$~sr$^{-1}$.

The generalization of the radio emissivity, $j_{\nu}$, to account for three CR populations,
each characterized by different power-law spectra and the inclusion of the
normalization parameter $g_{\rmn{CR}}$, following \cite{2010MNRAS.409..449P}, is
straight forward. We obtain
\begin{eqnarray}
j_{\nu,\rmn{extended}} & = &g_{\rmn{CR}} C(R) \rho_{\rmn{gas}}(R) \frac{\epsilon_{\rmn{B}}(R)}{\epsilon_{\rmn{B}}(R)+\epsilon_{\rmn{CMB}}} \nonumber \\
& \times & \Sigma_{i=1}^{3} \Delta_{i} A_{\nu,i} \left( \frac{\epsilon_{\rmn{B}}(R)}{\epsilon_{B_{\rmn{c}}}} \right)^{\frac{\alpha_{i}-2}{4}}  \, ,
\end{eqnarray}
where the sum is over the three CR spectral indexes $\alpha_{i}=(2.55,2.3,2.15)$
with the corresponding factors $\Delta_{i} = (0.767, 0.143, 0.0975)$ found by
\cite{2010MNRAS.409..449P}.


\section{Gamma-ray Emission}
\label{app:C}

The gamma-ray flux above an energy $E_{\gamma}$ is given by (e.g.,
\citealp{2010MNRAS.409..449P})
\begin{equation}
F_{\gamma} (>E_{\gamma}) = \frac{1}{4\pi D^{2}} \int dV  \lambda_{\gamma}(R)\, ,
\end{equation}
where the omnidirectional (i.e., integrated over the $4\pi$ solid angle)
gamma-ray emissivity above $E_{\gamma}$ is $ \lambda_{\gamma}(R)=A_{\gamma} C(R)
\rho_{\rmn{gas}}(R)$. The parameter $A_{\gamma}$ is \citep{2010MNRAS.409..449P}
\begin{eqnarray}
A_{\gamma} = g_{\rmn{CR}} D_{\gamma,\rmn{break}} \frac{4 m_{\pi^{0}} c}{3 m_{\rmn{p}}^{2}} \Sigma_{i=1}^{3} \Delta_{i} \frac{\sigma_{\rmn{pp},i}}{\alpha_{i} \delta_{i}} \left( \frac{m_{\rmn{p}}}{2 m_{\pi^{0}}} \right)^{\alpha_{i}} \times \nonumber \\
\times \left[ \beta_{x} \left( \frac{\alpha_{i}+1}{2\delta_{i}}, \frac{\alpha_{i}-1}{2\delta_{i}} \right) \right]_{x_{1}}^{x_{2}} \, ,
\end{eqnarray}
where $x_{j}=\{ 1 + [ m_{\pi^{0}}c^2/(2E_{\gamma,j})]^{2\sigma_{i}} \}$, $\left[ \beta_{x}(a,b) \right]_{x_1}^{x_2} =
\beta_{x_2}(a,b)-\beta_{x_1}(a,b)$ where $\beta$ denotes the incomplete
Beta-function \citep{1965hmfw.book.....A}, and
$\delta_{i}=0.14\alpha_{i}^{-1.6}+0.44$. The term $D_{\gamma,
  \rmn{break}}=D_{\gamma}(E_{\gamma},E_{\gamma,\rmn{break}})$ represents
diffusive CR losses due to escaping protons from the cluster at the equivalent
photon energy for the break $E_{\gamma, \rmn{break}}$ (see
\citealp{2010MNRAS.409..449P} for details). $A_{\gamma}$ is given in units of
cm$^3$~s$^{-1}$~g$^{-1}$.


\section{Observational Radio-to-X-ray Scaling relation and Luminosity Function}
\label{app:D}

For comparison with the observed 1.4~GHz radio-to-X-ray scaling relation, we
include almost all RHs of the sample by \cite{2011A&A...527A..99E}. 
We exclude RXCJ1314.4-2515 and Z7160 because they lack bolometric X-ray measurements. 
We add to our sample the clusters Ophiuchus, A2029 and A1835
\citep{2009A&A...499..371G}. The bolometric X-ray luminosities,
$L_{\rmn{X,bol}}$, of clusters hosting giant radio halos are taken from
\cite{2009A&A...507..661B}, while $L_{\rmn{X,bol}}$ values of cluster hosting
for mini-halos are taken from \cite{2002ApJ...567..716R},
\cite{Boehringer:1998vv} and ACCEPT. In the cases where measurement
uncertainties for X-ray and radio luminosities are not reported, we adopt a
10 per cent uncertainty. Our final RH sample has a median redshift of
$z\approx0.18$. In the left panel of Fig.~\ref{fig:PLSZ}, we show the
corresponding radio-to-X-ray scaling relation which is well fit by $\log_{10}
(L_{1.4~\rmn{GHz}}/h_{70}^{-2}~\rmn{erg}~\rmn{s}^{-1}~\rmn{Hz}^{-1}) = A +
B~\log_{10} (L_{\rmn{X,bol}}/h_{70}^{-2}~\rmn{erg}~\rmn{s}^{-1})$, with
$A=-37.204\pm1.838$, $B=1.512\pm0.041$, and scatter $\sigma_{yx} \approx
0.52$. Regarding the clusters with upper limits on the diffuse radio emission in
the sample of \cite{2011A&A...527A..99E}, we select only those 8 clusters with
ACCEPT measurements to obtain a homogeneous data set in $L_{\rmn{X,bol}}$ (note
that there are a number of clusters with upper limits on $L_{1.4~\rmn{GHz}}$ for
which there are only soft-band X-ray luminosities available).

\begin{figure} 
\centering
\includegraphics[width=0.48\textwidth]{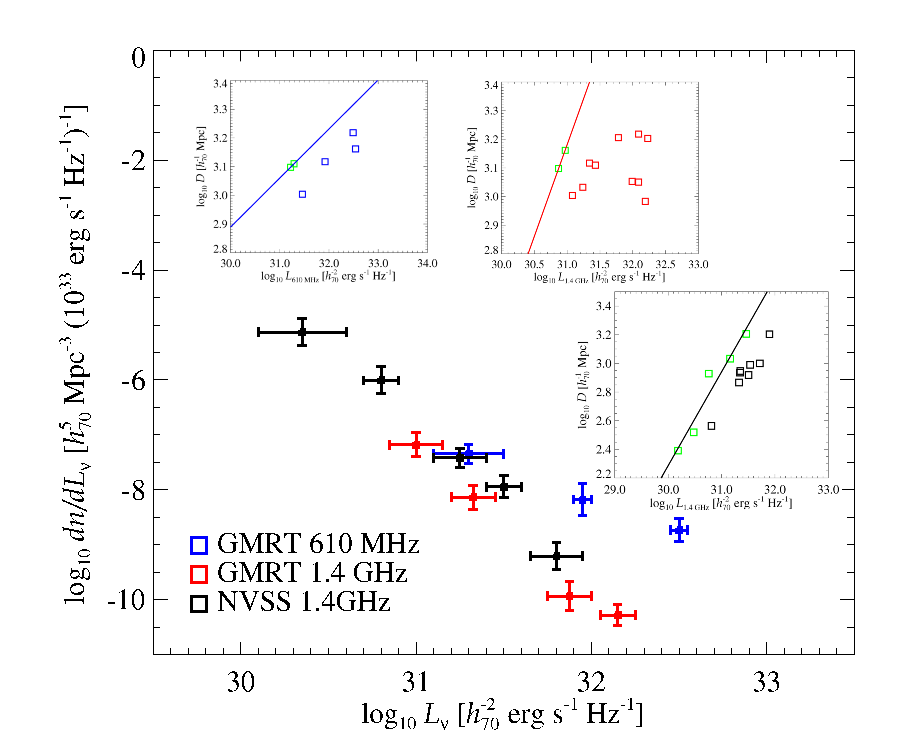}
\caption{RH luminosity function obtained from existent observations. The insets
  show the sample luminosity-distance distributions (see main text for details)
  where the solid line is the fit to the upper envelope population (indicated in
  green) employed to calculate the flux limit for the classical $V_{\rmn{max}}$
  estimator. The choice of the upper envelope population is somehow arbitrary,
  particularly in the GMRT cases due to the poor luminosity--distance
  distributions. The horizontal error bars represent the mass bins while the
  vertical error bars are Poissonian uncertainties.}
\label{fig:RLFobs}
\end{figure}

In Fig.~\ref{fig:RLFobs}, we make an attempt to construct an RLF from existing
X-ray flux-limited radio surveys. We consider the
\cite{1999NewA....4..141G} survey with NVSS at $1.4$~GHz and the
\cite{VenturiGMRT_1,VenturiGMRT_2} survey with GMRT at $610$~MHz by. We only
select RHs, i.e., we do not consider radio relics or other diffuse radio
emissions of unclear classification. The 1.4~GHz NVSS survey contains 13~RHs out
of 205 analyzed clusters while the 610~MHz GMRT survey contains 6 RHs out of the
34 observed. The sample finally analyzed by \cite{VenturiGMRT_1,VenturiGMRT_2}
is composed by 50 clusters and we can also build a corresponding RLF at 1.4~GHz
using the 12 present RHs. The fraction of radio-loud clusters is about 0.06, 0.18
and 0.24 for the NVSS 1.4~GHz, GMRT 610~MHz and GMRT 1.4~GHz sample,
respectively. The corresponding median redshift is 0.18, 0.26 and 0.25. We
calculate the RLF using the classical $V_{\rmn{max}}$ estimator (e.g.,
\citealp{1976ApJ...207..700F}) correcting for the sample incompleteness and
survey sky coverage. The most problematic aspect in obtaining these RLFs, apart
from the few available objects, is the calculation of a meaningful flux
limit. We obtain such a limit by fitting the upper envelope of the
luminosity-distance distribution of a given sample, as shown in the insets of
Fig.~\ref{fig:RLFobs}, following the procedure adopted by
\cite{2011arXiv1106.5494B}. Note that it is particularly ambiguous to calculate
a meaningful flux limit for the GMRT survey due to its poor luminosity-distance
distribution. We decide therefore to take the 1.4~GHz NVSS RLF as reference in
our comparisons with observation. However, we want to stress that several issues
can affect this result, such as the small sample size that impacts the resulting
flux limit, and the Malmquist and Eddington biases. Indeed, the very different
fraction of radio loud clusters obtained from different samples is one clear
indicator of the large uncertainty in the RLF. 

\end{appendix}

\label{lastpage}

\end{document}